\documentclass[11pt]{article}
\usepackage{titlesec}
\usepackage{hyperref}

\titleclass{\subsubsubsection}{straight}[\subsection]

\usepackage[utf8]{inputenc}
\usepackage{amsmath,amsfonts,amssymb,stackengine,graphicx}

\usepackage[utf8]{inputenc}
\newif\iffigs\figstrue

\usepackage[title]{appendix}
\usepackage{epsfig,latexsym}
\usepackage{hyperref}
\usepackage{amsmath}
\usepackage{verbatim}
\usepackage{color}
\usepackage{mathrsfs}
\usepackage{slashed}
\usepackage{amssymb}

\DeclareMathAlphabet{\mathpzc}{OT1}{pzc}{m}{it}

 \csname
@addtoreset\endcsname{equation}{section}

\def\gz0{\gamma^{0}}

\def\beq{\begin{equation}}
\def\eeq{\end{equation}}
\def\bea{\begin{eqnarray}}
\def\eea{\end{eqnarray}}
\def\ba{\begin{array}}
\def\ea{\end{array}}
\def\bec{\begin{center}}
\def\ec{\end{center}}
\def\ba{\begin{align}}
\def\ena{\end{align}}

\def\12{\frac{1}{2}}

\def\pr{\partial}

\newcounter{subsubsubsection}[subsubsection]
\renewcommand\thesubsubsubsection{\thesubsubsection.\arabic{subsubsubsection}}

\titleformat{\subsubsubsection}
  {\normalfont\normalsize\bfseries}{\thesubsubsubsection}{1em}{}
\titlespacing*{\subsubsubsection}
{0pt}{3.25ex plus 1ex minus .2ex}{1.5ex plus .2ex}

\makeatletter
\renewcommand\paragraph{\@startsection{paragraph}{5}{\z@}%
  {3.25ex \@plus1ex \@minus.2ex}%
  {-1em}%
  {\normalfont\normalsize\bfseries}}
\renewcommand\subparagraph{\@startsection{subparagraph}{6}{\parindent}%
  {3.25ex \@plus1ex \@minus .2ex}%
  {-1em}%
  {\normalfont\normalsize\bfseries}}
\def\toclevel@subsubsubsection{4}
\def\toclevel@paragraph{5}
\def\toclevel@paragraph{6}
\def\l@subsubsubsection{\@dottedtocline{4}{7em}{4em}}
\def\l@paragraph{\@dottedtocline{5}{10em}{5em}}
\def\l@subparagraph{\@dottedtocline{6}{14em}{6em}}
\makeatother

\begin{document}

\begin{flushright}
{\today}
\end{flushright}

\vspace{10pt}

\begin{center}


{\Large\sc Brane Profiles of Non--Supersymmetric Strings}\\


\vspace{25pt}
{\sc J.~Mourad${}^{\; a}$, S.~Raucci${}^{\; b}$  \ and \ A.~Sagnotti${}^{\; b}$\\[15pt]

${}^a$\sl\small APC, UMR 7164-CNRS, Universit\'e  Paris Cit\'e \\
10 rue Alice Domon et L\'eonie Duquet \\75205 Paris Cedex 13 \ FRANCE
\\ e-mail: {\small \it
mourad@apc.univ-paris7.fr}\vspace{10
pt}

{${}^b$\sl\small
Scuola Normale Superiore and INFN\\
Piazza dei Cavalieri, 7\\ 56126 Pisa \ ITALY \\
e-mail: {\small \it salvatore.raucci@sns.it, sagnotti@sns.it}}\vspace{10pt}
}

\vspace{40pt} {\sc\large Abstract}\end{center}
\noindent
We connect the indications of 2D CFT for branes in non--supersymmetric strings to actual spacetime profiles, taking the bulk tadpole potentials into account. We find exact solutions for the uncharged branes that spread in the internal intervals of Dudas--Mourad vacua for non--tachyonic orientifolds and heterotic strings. These solutions involve suitable dressings of the uncharged branes of the nine--dimensional tadpole--free theory. Similar exact results for uncharged branes that are transverse to the internal space, or for charged branes, appear more challenging.
Nevertheless, we identify their large--distance behavior, which is determined by linearized equations, and show that it is compatible with the CFT analysis in all expected cases. We also provide some hints on the leading back-reaction of the form-fields.

\setcounter{page}{1}

\pagebreak

\newpage
\tableofcontents
\newpage
\baselineskip=20pt
\section[Introduction and Summary]{\sc  Introduction and Summary}\label{sec:intro}

Allowing for the breaking of supersymmetry in a controllable fashion is a difficult challenge in String Theory~\cite{strings}, which cannot be foregone
in view of the sought connection with Particle Physics. A number of mechanisms have been explored, but while the resulting scenarios are enticing and often suggestive, a precise control of the back-reaction introduced by broken supersymmetry appears beyond all currently available techniques, which confine the analysis to low--energy effective descriptions. 

Supersymmetry breaking typically brings along tachyonic instabilities, which remain particularly challenging despite several attempts to clarify their fate (for a recent review, see~\cite{sen_zwiebach}), but there are three ten--dimensional string models~\cite{so1616,susy95,sugimoto} where this vexing problem is absent.
The first non--tachyonic ten--dimensional string is a variant of the supersymmetric heterotic models of~\cite{heterotic}, while the other two are orientifolds~\cite{orientifolds}. There is also a notable difference between the models of~\cite{so1616,susy95} and that of~\cite{sugimoto}, because supersymmetry is absent in the first two, but is present, albeit non--linearly realized~\cite{bsb,bsb_nonlinear}, in the latter. Yet, from the low--energy vantage point, as we shall review shortly, the three non--tachyonic models can be addressed along similar lines.

The emergence of a tadpole potential is the leading back-reaction induced by broken supersymmetry in all three non--tachyonic ten--dimensional strings. Its main consequence is that ten--dimensional Minkowski space is not a vacuum.
When the tadpole potential is taken into account, the most symmetric solutions are the Dudas--Mourad vacua introduced in~\cite{dm_vacuum}, which have the enticing feature of giving rise to an effective nine--dimensional Minkowski space. The main theme of the present paper are the brane profiles that emerge in these vacua.

In the Einstein frame, the relevant portion of the low--energy effective action for the three cases of interest involves the metric, the dilaton and a form field strength, and reads
 \bea \label{eq:einstein_frame_action}
{\cal S} &=& \frac{1}{2\kappa_{10}^2}\int d^{10}x \, \sqrt{-g}\left[\mathcal{R}\ - \ \frac{1}{2}\ (\partial\phi)^2\ - \ T \, e^{\,\gamma\,\phi}\ - \ \frac{e^{-2\,\beta_p\,\phi}}{2\,(p+2)!}\, {\cal H}_{p+2}^2
\right] \ .
\eea

For the heterotic SO(16) $\times$ SO(16) model,
\beq
\gamma \ = \ \frac{5}{2} \ ,
\eeq
and there are charged $p$-branes with $p=1,5$ and
\beq
\beta_p \ = \ \frac{3-p}{4} \ ,
\eeq
which are fundamental strings and NS5 branes.

For the two orientifold models,
\beq
\gamma \ = \ \frac{3}{2} \ ,
\eeq
and in the USp(32) model of~\cite{sugimoto} there are charged Dp-branes with $p=1,5$. On the other hand, for the type 0'B model of~\cite{susy95} the charged branes have $p=1,3,5$, and for $p=3$ the field strength satisfies the self--duality condition
\beq
{\cal H}_5 \ = \ \star\,{\cal H}_{5} \ .
\eeq
Moreover, for both orientifold models
\beq
\beta_p \ = \ \frac{p-3}{4}\ .
\eeq

In some of the ensuing analysis it will prove convenient to allow for generic values of the spacetime dimension ${\cal D}$, thus working with
 \bea
{\cal S}_E &=& \frac{1}{2\kappa_{\cal D}^2}\int d^{\cal D}x \, \sqrt{-g}\left[\mathcal{R}\ - \ \frac{4}{{\cal D}-2}\ (\partial\phi)^2\ - \ T \, e^{\,\gamma\,\phi}\ - \ \frac{e^{-2\,{\beta}_{{\cal D},p}\,\phi}}{2\,(p+2)!}\, {\cal H}_{p+2}^2  \
\right] \ , \label{eqs4}
\eea
although this extension does not concern directly the critical strings of interest for this paper. Our notation for $\gamma $ and $\beta_{{\cal D},p}$ is spelled out in detail in~\cite{mrs24_1}. We shall mostly work in ten dimensions and, for brevity, we shall mostly leave the ${\cal D}$ subscript in $\beta_{{\cal D},p}$ implicit.

In general, the portions of the Einstein--frame equations concerning the metric tensor, a $(p+1)$-form gauge field and the dilaton read
\bea
\mathcal{R}_{MN} \,-\, \frac{1}{2}\ g_{MN}\, \mathcal{R} &=& \!\!\frac{4}{{\cal D}-2}\  \pr_M\phi\, \pr_N\phi\, + \, \frac{e^{\,-\,2\,\beta_{{\cal D},p}\,\phi} }{2(p+1)!}\,  \left({\cal H}_{p+2}^2\right)_{M N}  \nonumber\\
&-& \!\!\frac{1}{2}\,g_{MN}\Big[\frac{4\,(\pr\phi)^2}{{\cal D}-2}+ \frac{e^{\,-\,2\,\beta_{{\cal D},p}\,\phi}}{2(p+2)!}\,{\cal H}_{p+2}^2\,+\, T \, e^{\gamma\,\phi}\Big] \ ,  \nonumber
\\
\frac{8}{{\cal D}-2} \ \Box\phi &=& \,-\,\frac{\beta_{{\cal D},p}\, e^{\,-\,2\,\beta_{{\cal D},p}\,\phi}}{(p+2)!} \ {\cal H}_{p+2}^2 \, +\, T \, \gamma \, e^{\gamma \,\phi}   \ , \nonumber
\\
d\Big(e^{\,-\,2\,\beta_{{\cal D},p}\,\phi}\ \star\,{\cal H}_{p+2}\Big) &=& 0  \ . \label{eqsbeta}
\eea
Equivalently, one can work with
\bea
{\cal R}_{MN}  &=& \frac{4}{{\cal D}-2}\  \pr_M\phi\, \pr_N\phi\ + \ \frac{1}{2(p+1)!}\ e^{\,-\,2\,\beta_{{\cal D},p}\,\phi} \ \left({\cal H}_{p+2}^2\right)_{M N}\nonumber\\
&+& g_{MN}\left[\,- \, \frac{(p+1)\, e^{\,-\,2\,\beta_{{\cal D},p}\,\phi}}{2({\cal D}-2)(p+2)!}\, {\cal H}_{p+2}^2 \  +\ \frac{T}{{\cal D}-2}\, e^{\gamma\,\phi}\right] \ . \label{eqsnotlagbeta}
\eea

The plan of this paper is the following. In Section~\ref{sec:sec2} we summarize some key properties of brane profiles in the absence of tadpole potentials, following~\cite{mrs24_1}. We begin by recalling the general harmonic--gauge setup, where the exact solutions take a simpler form, and develop it further in Section~\ref{sec:charged_branes}, which is devoted to charged branes. In Section~\ref{sec:uncharged_branes} we turn to uncharged branes, and we also recast their description in the isotropic gauge, where the solutions are more involved but afford a simpler physical interpretation. This presentation will play an important role in the following sections. In Section~\ref{sec:branes_with_bulk_tadpoles} we turn to the main topic of this paper, and address the general setup for describing branes whose profiles depend on two independent variables, respect the proper isometries and approach the vacuum asymptotically at large distances from the core. This setup is instrumental to address the branes that emerge, in nine dimensions, from the compactifications on the Dudas--Mourad vacua of~\cite{dm_vacuum}. In Section~\ref{sec:vacuum_solution} we also recall the key properties of those vacua, with some emphasis on their singular behavior near the endpoints, before setting up the full non--linear system for the resulting brane profiles in Sections~\ref{sec:brane_profiles} and~\ref{sec:brane_equations}. In Section~\ref{sec:exact_solutions} we show that, after a proper dressing, nine--dimensional uncharged branes yield exact solutions in the presence of a dilaton tadpole. The case of charged branes appears more challenging, and we have not succeeded to find exact solutions for them. However, the large--distance behavior can be captured by the linearized system of equations that, while still complicated, can be solved exactly. The asymptotic solutions are discussed in detail in Section~\ref{sec:linearized_analysis}, where we show that all the Dp-branes that emerged from the CFT analysis in~\cite{dms_cft} can be recovered in this fashion. This Section also completes the analysis in~\cite{ms23_1}, while also correcting some statements contained in Section~2 of that paper concerning vectors and form fields. Section~\ref{sec:Conclusions} contains our Conclusions and, finally, Appendix~\ref{app:brane_details} collects a number of details on the derivation of the complete non--linear system.

\section[Branes Without Tadpole Potentials]{\sc Branes Without Tadpole Potentials} \label{sec:sec2}

In this section we summarize some results of~\cite{mrs24_1} that play a role in the present discussion. They concern the class of ${\cal D}$--dimensional backgrounds
\bea
ds^{\,2} &=& e^{2A(r)}\, \eta_{\mu\nu}\, dx^\mu\,dx^\nu \ + \ e^{2B(r)}\,dr^2\ + \ e^{2C(r)}\, \ell^2 \,d\Omega_{{\cal D}-p-2}^2  \ , \nonumber \\
\phi &=& \phi(r) \ , \nonumber \\
{\cal H}_{p+2} &=& H_{p+2}\ e^{\,2\,\beta_p\phi + B +(p+1)A-({\cal D}-p-2)C}\  dx^0 \wedge \ldots \wedge dx^p \wedge dr \ , \label{metric_sym}
\eea
where $\mu=0,\ldots,p$, $d\Omega_{{\cal D}-p-2}^2$ is the metric on a round sphere of unit radius and $\ell$ is a length scale. The isometries of this class of backgrounds combine the spacetime Poincar\'e symmetry with spherical symmetry in the internal space, and ${\cal H}_{p+2}$ is a form-field strength compatible with these isometries that solves the corresponding Bianchi identities and equations of motion. It describes a generic ``electric'' profile characterized by the overall constant $H_{p+2}$, whose presence is instrumental to support charged branes.

The class of metric in eqs.~\eqref{metric_sym} retains some residual gauge symmetry related to presence of $B(r)$, which could clearly be absorbed redefining the $r$ variable. However, the residual symmetry was essential to derive the exact solutions of~\cite{dm_vacuum}, and it will also simplify matters here, where we shall eventually make two different gauge choices. The first is the harmonic gauge,
\beq
B \ = \ (p+1)A \ + \ ({\cal D}-p-2) C \ ,
\eeq
which eliminates some non--linear contributions to the background equations, leading to relatively simple exact solutions, but at the price of giving rise to unfamiliar asymptotic limits. The second choice is the isotropic gauge,
\beq
C \ = \ B \ + \ \log\,r \ ,
\eeq
whose indications are physically more transparent, in particular for what concerns the asymptotically flat limit, which corresponds to
\beq \label{eq:asymptotics_tadpole-free_branes}
A(r) \ \underset{r\to\infty}{\to} \ 0 \ , \qquad  B(r) \ \underset{r\to\infty}{\to} \ 0 \ ,
\eeq
but where the solutions take a more complicated form, which can be conveniently reached starting from the simpler harmonic--gauge expressions.

The isometries of eqs.~\eqref{metric_sym}, together with the limiting behavior in eq.~\eqref{eq:asymptotics_tadpole-free_branes}, characterize general $p$-brane solutions in a flat background.

\subsection{\sc Charged Branes} \label{sec:charged_branes}

Let us begin our discussion by recalling some basic results on charged branes in the language of~\cite{mrs24_1}.
In the harmonic gauge, one can derive analytically six classes of exact solutions of this type. They are distinguished by the possible sign choices for a pair of integration constants ${\cal E}_x$ and ${\cal E}_y$ compatible with the Hamiltonian constraint, which links them to another integration constant $z_1$, according to
\beq \label{eq:deformed_charged_z1}
z_1^2 \ =  \ \frac{1}{4(p+1)}\left[\frac{({\cal D}-p-2)}{4({\cal D}-2)}\,\Delta \, {\cal E}_x \ - \ ({\cal D}-p-3) \, {\cal E}_y\right] \ ,
\eeq
where
\beq
\Delta \ = \ 4({\cal D}-p-3)(p\,+\,1) \ +\  ({\cal D}-2)^2 \beta_p{}^2  \ . \label{Delta}
\eeq
Letting
\beq
{\cal F}\left(\mathcal{E}, r \right)\ = \ \begin{cases}
    \frac{1}{\sqrt{\mathcal{E}}}\sinh{\left(\sqrt{\mathcal{E}} \, r\right)} \qquad & \text{if   } \mathcal{E}>0 \ , \\
    r & \text{if   } \mathcal{E}=0 \ , \label{cases_F} \\
    \frac{1}{\sqrt{|\mathcal{E}|}}\sin{\left(\sqrt{|\mathcal{E}|} \, r\right)} & \text{if   } \mathcal{E}<0 \ ,
\end{cases}
\eeq
these backgrounds are described by
\bea \label{eq:deformation_backgroun_general}
ds^2 &=& \left|\frac{{\cal F}\left({\cal E}_y, r+r_1 \right)}{{\cal F}\left({\cal E}_y, r_1 \right)}\right|^{-\frac{8({\cal D}-p-3)}{\Delta}} e^{\frac{8({\cal D}-2)\beta_p}{\Delta} z_1 r }  dx_{p+1}^2 \nonumber \\  &+& \left|\frac{{\cal F}\left({\cal E}_y, r+r_1 \right)}{{\cal F}\left({\cal E}_y, r_1 \right)}\right|^{\frac{8(p+1)}{\Delta}} e^{-\frac{8({\cal D}-2)(p+1)\beta_p}{({\cal D}-p-3)\Delta} z_1 r } \left(\frac{dr^2}{\left|\frac{({\cal D}-p-3)}{\ell } \, {\cal F}\left({\cal E}_x, r \right) \right|^\frac{2({\cal D}-p-2)}{{\cal D}-p-3}} \right.  \nonumber \\
&+& \left. \ \frac{\ell^2\ d\Omega_{{\cal D}-p-2}^2}{\left|\frac{({\cal D}-p-3)}{\ell } \, {\cal F}\left({\cal E}_x, r \right) \right|^\frac{2}{{\cal D}-p-3}} \right) \ , \nonumber \\
e^\phi &=&  e^{\phi_0}\ e^{-\frac{4({\cal D}-2)(p+1)}{\Delta} z_1 r}  \left|\frac{{\cal F}\left({\cal E}_y, r+r_1 \right)}{{\cal F}\left({\cal E}_y,r_1 \right)}\right|^{-\frac{\left({\cal D}-2\right)^2\beta_p}{\Delta}} \ ,  \\
{\cal H}_{p+2} &=& - \ \epsilon \  \sqrt{\frac{8({\cal D}-2)}{\Delta}} \ e^{\beta_p \phi_0} \ \left|{\cal F}\left({\cal E}_y, r_1 \right)\right|^{-1}   \left|\frac{{\cal F}\left({\cal E}_y, r+r_1 \right)}{{\cal F}\left({\cal E}_y, r_1 \right)}\right|^{-2} \ dx^0 \wedge \ldots \wedge dx^p \wedge dr \ ,\nonumber
\eea
where $\ell$ is again an arbitrary length scale.
These backgrounds thus involve four real parameters, $\phi_0$, two additional ones out of ${\cal E}_x$, ${\cal E}_y$ and $z_1$, taking into account the Hamiltonian constraint~\eqref{eq:deformed_charged_z1}, and the positive constant $r_1$, together with a sign choice characterized by $\epsilon=\pm1$.
In eqs.~\eqref{eq:deformation_backgroun_general}, the range of $r$ is delimited by the zeroes of ${\cal F}\left({\cal E}_y, r+r_1 \right)$ and ${\cal F}\left({\cal E}_x, r \right)$, and if one of the endpoints is a zero of ${\cal F}\left({\cal E}_x, r \right)$, the corresponding solution is asymptotically flat. In those cases, the brane tension and charge can be recovered expanding the background near the end where the zero of ${\cal F}\left({\cal E}_x, r \right)$ lies, and one finds~\footnote{We are considering a dilaton dressing along the line of what happens for BPS branes, namely $e^{\beta_p \phi} {\cal T}_p$, without specifying the dependence of $\beta_p$ on the brane dimension $p$.}
\bea \label{eq:charged_tension_charge}
{\cal T}_p &=& e^{-\beta_p \phi_0} \frac{8({\cal D}-2)}{\Delta}\frac{\Omega_{{\cal D}-p-2}}{2\kappa_{\cal D}^2} \, \ell^{ \,{\cal D}-p-2 }\left[\frac{{\cal F}'\left({\cal E}_y, r_1\right)}{{\cal F}\left({\cal E}_y, r_1\right)} \ - \ \frac{{\cal D}-2}{{\cal D}-p-3}\ \beta_p z_1\right] \ , \nonumber \\
Q_p &=& \epsilon \  e^{-\beta_p \phi_0} \sqrt{\frac{8({\cal D}-2)}{\Delta}}\frac{\Omega_{{\cal D}-p-2}}{2\kappa_{\cal D}^2}\,\ell^{ \,{\cal D}-p-2 }\left|{\cal F}\left({\cal E}_y, r_1\right)\right|^{-1} \ ,
\eea
where $\Omega$ denotes the area of a unit sphere.

The BPS branes of String Theory are recovered for the special values ${\cal E}_x={\cal E}_y=z_1=0$, and their tensions and charges, which are identical up to a relative sign, are both determined by $r_1$. In the general case $\phi_0$ determines the asymptotic value of the dilaton, $z_1$ contributes to $\phi'(0)$, the asymptotic value of its radial derivative at large distances from the brane, and tension and charge are determined by the remaining independent parameters.

\subsection{\sc Uncharged Branes}
\label{sec:uncharged_branes}

Uncharged branes can be obtained from eqs.~\eqref{eq:deformation_backgroun_general} by sending $r_1\to\infty$, when ${\cal E}_x>0$ and ${\cal E}_y\geq0$.
In this limit, the background becomes
\bea
ds^2 &=&  \ e^{-\frac{2\,r}{R}}dx_{p+1}^2 \nonumber \ + \ e^{\frac{2(p+1)\,r}{\left({\cal D}-p-3\right)R}} \left[\frac{({\cal D}-p-3)\sigma}{\ell} \,\sinh{\left(\frac{r}{\sigma}\right)}\right]^{-\,\frac{2\left({\cal D}-p-2\right)}{{\cal D}-p-3}} dr^2 \nonumber \\ &+& e^{\frac{2(p+1)\,r}{\left({\cal D}-p-3\right)R}}\left[ \frac{({\cal D}-p-3)\sigma}{\ell} \,\sinh{\left(\frac{r}{\sigma}\right)}  \right]^{-\,\frac{2}{{\cal D}-p-3}} \ell^2 \, d\Omega_{{\cal D}-p-2}^2 \ , \nonumber\\
\phi & = & \phi_0 \ + \ \phi_1 r \ , \label{eq:uncharged_harmonic}
\eea
with
\bea
\sigma & = & \frac{1}{\sqrt{{\cal E}_x}} \ , \qquad \frac{1}{R} \ = \  \frac{4({\cal D}-p-3)}{\Delta}\,\sqrt{{\cal E}_y}\ - \ \frac{4({\cal D}-2)\beta_p}{\Delta} \, z_1 \ ,  \nonumber \\
\phi_1 &=& - \ \frac{({\cal D}-2)^2\beta_p}{\Delta}\, \sqrt{{\cal E}_y} \  - \ \frac{4({\cal D}-2)(p+1)}{\Delta}\, z_1 \ ,
\eea
where $\Delta$ is defined in eq.~\eqref{Delta}.
The Hamiltonian constraint of eq.~\eqref{eq:deformed_charged_z1} can then be cast in the form
\beq
\frac{1}{R^2} \ \frac{(p+1)({\cal D}-2)}{{\cal D}-p-3} \ + \ \frac{4\,\phi_1^2}{{\cal D}-2} \ = \ \frac{1}{\sigma^2} \ \frac{{\cal D}-p-2}{{\cal D}-p-3} \ , \label{ham_red}
\eeq
so that the background has three real parameters, which determine the asymptotic value of the dilaton, $\phi'(0)$ and the tension of the brane.

From the harmonic--gauge profiles of eqs.~\eqref{eq:uncharged_harmonic}, in~\cite{mrs24_1} we deduced corresponding expressions for metric and dilaton in the isotropic gauge, which read
\bea \label{eq:uncharged_isotropic}
ds^2 &=& \left[\frac{1+v(r)}{1-v(r)}\right]^{\frac{-2\, \sigma}{R}} dx_{p+1}^2 \nonumber \\
&+& \left[\frac{1+v(r)}{1-v(r)}\right]^{\frac{2\,\sigma \left(p+1\right)}{R\left({\cal D}-p-3\right)}} \left[1\,-\,v^2(r)\right]^{\frac{2}{{\cal D}-p-3}} \left(dr^2+ r^2  d\Omega_{{\cal D}-p-2}^2\right) \
, \nonumber \\
\phi &=& \phi_0 \ + \ \phi_1 \,\sigma \log\left[\frac{1\,+\,v(r)}{1\,-\,v(r)}\right] \ ,
\eea
where
\beq \label{eq:vrho}
v(r)\ = \ \frac{\ell}{2({\cal D}-p-3) \sigma} \left(\frac{\ell}{r}\right)^{{\cal D}-p-3} \  .
\eeq

As explained in~\cite{mrs24_1}, the limiting behavior of the metric as $r\to \infty$ determines the tension
\beq
{\cal T}_p \ = \ e^{-\beta_p \phi_0} \,\frac{ ({\cal D}-2) \, \Omega_{{\cal D}-p-2} }{ ({\cal D}-p-3) \, \kappa_{\cal D}^2 }\frac{\,\ell^{ \,{\cal D}-p-2 }}{R} \ ,
\eeq
which coincides with the limit of eq.~\eqref{eq:charged_tension_charge} as $r_1\to\infty$.
The tension is positive (negative) for positive (negative) values of $R$.

If the dilaton profile is constant or absent altogether, $\phi_1$ vanishes and the Hamiltonian constraint reduces to
\beq
R \ = \ \pm  \ \sqrt{\frac{(p+1)({\cal D}-2)}{{\cal D}-p-2}} \  \sigma  \ ,
\eeq
so that the two integration constants $R$ and $\sigma$ are proportional to one another. This special class of solutions, which have also the virtue of not including any strong--couplings regions, will play a central role even when the tadpole potentials of non--supersymmetric strings are present.

\section[Branes With Bulk Tadpoles]{\sc Branes With Bulk Tadpoles} \label{sec:branes_with_bulk_tadpoles}

As we stressed in the previous section, the charged and uncharged brane backgrounds of eqs.~\eqref{eq:deformation_backgroun_general} and~\eqref{eq:uncharged_harmonic} approach the vacuum, flat space in that case, at large distances from the singularity.
In a similar fashion, the gravitational backgrounds for branes in the presence of tadpole potentials should reduce, far from the localized objects, to the corresponding vacua. In the presence of bulk tadpoles, the most symmetric solutions of this type are the Dudas--Mourad vacua, which describe the spontaneous compactification of the ten--dimensional theory in an interval~\footnote{These solutions have cosmological counterparts that will be addressed, from this vantage point, in a different publication.}. The most relevant examples, with $\gamma=\frac{3}{2}$ and $\gamma=\frac{5}{2}$ in ten dimensions, were originally found in~\cite{dm_vacuum}. Generalizations to arbitrary values of $\gamma$ and ${\cal D}$ were then presented in~\cite{Russo:2004ym,Basile:2022ypo}. In the following, we shall concentrate on the ten--dimensional case, which is of direct relevance for String Theory. As we shall see, the uncharged branes of Section~\ref{sec:uncharged_branes} can be used to build exact solutions in these vacua. In this and the following section, we shall also explore the embedding of charged branes in these vacua, solving the corresponding linearized systems and examining the leading back-reactions of the form fields, showing that they are generally under control, up to a couple of instances where the conditions hold only slightly away from the endpoint singularities, where however the whole low-energy effective theory becomes problematic.

\subsection{\sc Vacuum Solutions With Tadpoles}
\label{sec:vacuum_solution}

The vacuum solutions can be captured working with a conformally flat metric and a dilaton profile of the type
\bea
ds^2  \ = \  e^{2\,\Omega(z)} \left(dx_{(9)}^2 \ + \ dz^2 \right) \ , \qquad \phi \ = \  \phi(z)  \ ,
\eea
and vanishing form fields
\beq
{\cal H}_{p+2} \ = \ 0 \ ,
\eeq
where the values of $p$ refer to all allowed form field strengths ($p=1,5$ for the USp(32) theory of~\cite{sugimoto} and for the  SO(16) $\times$ SO(16) theory of~\cite{so1616}, and $p=1,3,5$ for the 0'B theory of~\cite{susy95}).

The conformal factor $\Omega$ and the dilaton profile $\phi$ for the vacuum satisfy the equations
\bea \label{eq:background_eom}
\Omega'' \ +\  8 \left(\Omega'\right)^2 \ + \ \frac{T}{8} \ e^{2\Omega+\gamma\phi} &=& 0 \, \nonumber \\
72\left(\Omega'\right)^2\ -\ \frac{1}{2}\left(\phi'\right)^2 \ + \ T \  e^{2\Omega+\gamma\phi} &=&0 \ , \nonumber\\
\phi'' \ +\ 8\Omega'\phi'\ -\ T\gamma\  e^{2\Omega+\gamma\phi} &=& 0 \ ,
\eea
which follow from the Einstein-frame action of eq.~\eqref{eq:einstein_frame_action}.
The conformal variable $z$ has a finite range, and its ends, which we shall take to lie at $z=0 $ and $z=z_m$, host curvature singularities.

In detail, near the endpoints the orientifold background, with $\gamma=\frac{3}{2}$, behaves as
\bea
\Omega &\sim& \frac{1}{8}\log \left(\frac{z}{z_m}\right)  \ , \qquad \Omega\ \sim\ \frac{1}{8}\log \left(1 \ - \ \frac{z}{z_m}\right) \ , \nonumber \\
\phi &\sim& \frac{3}{2}\log \left(\frac{z}{z_m}\right)  \ , \qquad \phi\ \sim\ - \ \frac{3}{2}\log \left(1 \ - \ \frac{z}{z_m}\right) \ , \label{omegaphi_orientifold}
\eea
so that the solution is weakly coupled as $z\to0$ and strongly coupled as $z\to z_m$. Still, the tadpole potential, which gives rise to the finite range of $z$, leaves no imprints on the asymptotic behavior at both ends, which is surprisingly dominated by the free theory~\cite{ms21_2}.
In contrast, for the heterotic SO(16) $\times$ SO(16) background,
\bea
\Omega &\sim& \frac{1}{24}\log \left(\frac{z}{z_m}\right)  \ , \qquad \Omega\ \sim\ \frac{1}{8}\log \left(1 \ - \ \frac{z}{z_m}\right) \ , \nonumber \\
\phi &\sim& - \ \frac{5}{6}\log \left(\frac{z}{z_m}\right)  \ , \qquad \phi\ \sim\  \frac{3}{2}\log \left(1 \ - \ \frac{z}{z_m}\right) \ ,\label{omegaphi_heterotic}
\eea
so that the solution is strongly coupled as $z\to 0$ and weakly coupled as $z\to z_m$, but the correspondence with the free theory only holds at the weak--coupling end. At the strongly coupled end, the curvature singularity is akin to that encountered for D8 branes in the massive type IIA theory.

The perturbative stability of these vacua was first investigated in~\cite{bms_18}, where scalar perturbations were linked to formally positive--definite Schr\"odinger Hamiltonians. These, however, are singular at the ends, and a closer scrutiny of the allowed self--adjoint boundary conditions showed~\cite{ms23_1} that a single stable condition exists in the orientifold case, which also guarantees the presence of a massless graviton mode accompanied by massive scalar excitations. For the heterotic case one can carry out a similar analysis, relying however on the more general class of Schr\"odinger potentials discussed in~\cite{ms23_2}, which approximate very closely the actual ones for the perturbations. The end result is essentially identical to what is discussed there for the graviton--dilaton sector. There are stable boundary conditions leading to massless a graviton, but also others that lead to a purely massive gravity spectrum.

\subsection{\sc Brane Profiles} \label{sec:brane_profiles}

One can describe brane profiles in these backgrounds, generalizing the preceding ansatz in order to allow for the proper isometries. To begin with, one can distinguish, among the nine spacetime coordinates, $p+1$ of them parametrizing the world-volume of the brane, which we denote by $x$, and $8-p$ remaining ones, transverse to the brane, which we denote by $y^i$. The brane profiles of interest are thus generally of the type\footnote{Metrics with these isometries were first considered by Weyl in the four-dimensional case~\cite{Weyl:1917gp}, and then extended to higher dimensions in~\cite{Charmousis:2003wm}.}
\bea \label{eq:ansatz_2variables1}
ds^2 &=& e^{2A(z,r)}dx_{p+1}^2 \ + \ e^{2B(z,r)} \, dy^i\,dy^i \ + \ e^{2D(z,r)} dz^2 \ , \nonumber \\
\phi &=& \phi(z) \ + \ \varphi(z,r) \ , \qquad
{\cal H}_{p+2} \ = \ {\cal H}_{p+2}(z,r) \ , \qquad  {\cal H}_{p+3} \ = \  {\cal H}_{p+3}(z,r) \ ,
\eea
where $\phi(z)$ is the background dilaton profile and $r^2 = y^i y^i $, so that
\beq
dy^i\,dy^i \ = \ dr^2 \ + \ r^2\, d\,\Omega^2_{7-p}  \ .
\eeq
The embedding in the preceding vacua is granted if, as $r \to \infty$,
\beq
A(z,r)\  \sim \ \Omega(z) \ , \qquad B(z,r) \ \sim \ \Omega(z) \ , \qquad D(z,r) \ \sim \ \Omega(z) \ ,
\eeq
while $\varphi$, ${\cal H}_{p+2}$ and ${\cal H}_{p+3}$ tend to zero.
This setup is useful to describe extended objects perturbatively, which can suffice to identify their tensions and charges. However, it can be also regarded as a gauge--fixed version of a more general ansatz, which can potentially simplify the analysis of the field equations, along the lines of what was done in~\cite{dm_vacuum}. The more general ansatz relies on the introduction of two more functions, and is described in detail in Appendix~\ref{app:brane_details}.

In eqs.~\eqref{eq:ansatz_2variables1}, we considered two types of form field strengths. They are both allowed by the isometries of the background, and correspond to $p$- or $(p+1)$-branes that are either localized or wrapped along the $z$ direction.

A moving basis for the metric of eqs.~\eqref{eq:ansatz_2variables1} is
\bea
\theta^\mu\ = \ e^A\, dx^\mu\,,\qquad \theta^r\ = \ e^B\, dr \,, \qquad \theta^z\ = \ e^D \,dz\,,\qquad \theta^a\ = \ e^B \, r \, \tilde{\theta}^a\, ,
\eea
where the one-forms $\tilde{\theta}^a$ refer to the unit sphere,
with $\mu=0,\dots p$, $a=1,\dots 7-p$. Form profiles compatible with the symmetries of the background can be built combining $dx^0 \wedge\ldots\wedge dx^p$ with $dr$ and/or $dz$, depending on whether one is considering a $p$-brane,
\beq \label{eq:two_variable_ansatz_p_brane}
{\cal H}_{p+2} \ = \ H(z,r) \,\theta^r \, \theta^0 \ldots \theta^p \ + \ \widetilde{H}(z,r) \,\theta^z \, \theta^0 \ldots \theta^p \ ,
\eeq
or a $(p+1)$-brane whose world-volume is also extended along the $z$ direction,
\beq \label{eq:two_variable_ansatz_p+1_brane}
{\cal H}_{p+3} \ = \ h(z,r) \, \theta^0 \dots \theta^p  \,\theta^r\, \theta^z  \ .
\eeq
In special cases, when one is a NS brane and the other is type-IIA brane, one could even consider the two field strengths simultaneously, but this option does not concern ten--dimensional non--supersymmetric non--tachyonic strings, which are the main target of our analysis.

\subsection{\sc Brane Equations}
\label{sec:brane_equations}

Let us begin by considering the Bianchi identities and the equations of motion for the $(p+2)$-form field strength,
\beq
d\, {\cal H}_{p+2} \ = \ 0 \ , \qquad d\,\left(e^{-2\beta_p\phi} \star {\cal H}_{p+2}\right) \ = \  0 \ ,
\eeq
where ${\cal H}_{p+2}$ is given by eq.~\eqref{eq:two_variable_ansatz_p_brane}.
These equations translate into a pair of first--order equations linking $H$ and $\widetilde{H}$:
\bea
\left[H \ e^{(p+1)A+B}\right]_z  &=& \left[\widetilde{H} \ e^{(p+1)A+D}\right]_r \ , \nonumber \\
\left[\widetilde{H} \ r^{7-p} \ e^{-2\beta_p\phi + (8-p)B} \right]_z  &=& - \ \left[ H  \ r^{7-p} \  e^{-2\beta_p\phi + (7-p)B + D} \right]_r \ ,
\eea
where $[ \ ]_z$ is a shorthand for a derivative that we shall frequently use in the following.
One can solve the Bianchi identities by introducing the potential
\beq
{\cal B}_{p+1} \ = \ e^{-(p+1)A}  \ {\cal B}(z,r) \, \theta^0 \ldots \theta^p \ ,
\eeq
so that $H(z,r)$ and $\widetilde{H}(z,r)$ can be expressed in terms of ${\cal B}$ as
\bea \label{eq:H_and_H_tilde}
H & = &   e^{-(p+1)A-B} \ {\cal B}_r \ , \nonumber \\
\widetilde{H} &=&   e^{-(p+1)A-D} \ {\cal B}_z \ .
\eea
In terms of ${\cal B}$, the form equation then reads
\beq \label{eq:2_var_form}
\left( e^{2G} \ {\cal B}_z \right)_z\ = \   - \left(e^{2(G -B  + D)} \ {\cal B}_r  \right)_r \ .
\eeq
where
\beq
2G \ = \ -2\beta_p\phi -(p+1)A +  (8-p)B + (7-p)\log{r} - D \ .
\eeq

The second option of interest is the $(p+3)$-form field strength of eq.~\eqref{eq:two_variable_ansatz_p+1_brane}, for which the Bianchi identities are automatically satisfied and the equations of motion are simply solved by
\beq \label{eq:p+3_form}
h(z,r) \ = \ \frac{h_0}{r^{7-p}} \ e^{2\beta_{p+1}\phi - (7-p)B} \ ,
\eeq
where $h_0$ is a constant.

Let us now turn to the dilaton and metric equations.
Including both types of field strengths, the dilaton equation reads
\bea
  \Box\phi & = & T \, \gamma \,  e^{\gamma\phi} \ + \ \beta_p \left[e^{-2B} {\cal B}_r^2 \ + \  e^{-2D} {\cal B}_z^2 \right] \, e^{-2\beta_p\phi-2(p+1)A} \nonumber \\
 &+&  h_0^2 \ \beta_{p+1}  \, e^{2 \beta_{p+1}\phi - 2(7-p)(B + \log r)} \ .
\eea
In writing explicitly the d'Alembert operator $\Box \phi$, it is convenient to let
\beq
F^{\pm} \ = \ (p+1)A \ \mp\  B\ +\ (7-p)(B+\log{r}) \ \pm\  D \ ,
\eeq
so that the end result reads
\bea \label{eq:2_var_dilaton}
& &  \!\!\!\!\!\!\!\!   e^{-2B}\left( \phi_{rr}+\phi_r F^+_r\right) + e^{-2D}\left( \phi_{zz} +  \phi_z   F_z^-  \right) \ = \ T \, \gamma \,  e^{\gamma\phi}   \nonumber \\
    &+&  \!\!\! \beta_p \left(e^{-2B} {\cal B}_r^2 \ + \  e^{-2D}  {\cal B}_z^2 \right) \, e^{-2\beta_p\phi-2(p+1)A} +  \frac{h_0^2 \ \beta_{p+1}}{ r^{2(7-p)}} \,   e^{2\,\beta_{p+1}\,\phi\,-\,2\,(7-p)\,B} .
\eea
Finally, the metric equations are
\bea
\mathcal{G}_{MN} &=& \frac{1}{2}\  \pr_M\phi\, \pr_N\phi\, + \, \frac{e^{\,-\,2\,\beta_p\,\phi} }{2(p+1)!}\,  \left({\cal H}_{p+2}^2\right)_{M N} \, + \, \frac{e^{\,-\,2\,\beta_{p+1}\,\phi} }{2(p+2)!}\,  \left({\cal H}_{p+3}^2\right)_{M N}  \nonumber\\
&-& \frac{1}{2}\,g_{MN}\left[\frac{(\pr\phi)^2}{2}+ \frac{e^{\,-\,2\,\beta_p\,\phi}}{2(p+2)!}\,{\cal H}_{p+2}^2\,+ \frac{e^{\,-\,2\,\beta_{p+1}\,\phi}}{2(p+3)!}\,{\cal H}_{p+3}^2\,+\, V(\phi)\right] \ ,
\eea
and their different components read
\bea
\mu\nu&:& e^{-2B} \Big[ 2 \left(A-B- F^+\right)_{rr} +2 A_r F^+_r -  (p+1)A_r^2+ B_r^2 -(7-p)\left(B_r+\frac{1}{r}\right)^2\nonumber  \\
    &-&D_r^2 - (F^+_r)^2 \Big]  \ + \ \frac{e^{-2B}}{r^2} \, \tilde{R}+ e^{-2D}\Big[ 2 (A-D-F^- )_{zz} +2 A_z F_z^- - (p+1)A_z^2  \nonumber \\
            &-& B_z^2-(7-p)B_z^2 + D_z^2-(F_z^-)^2 \Big] \nonumber \\
                &=& \frac{1}{2}\, \left[e^{-2B} {\cal B}_r^2 \, + \,  e^{-2D} {\cal B}_z^2 \right] e^{\,-\,2\,\beta_p\,\phi\, -\,2\,(p+1)\,A} \, + \, \frac{h_0^2}{2 \, r^{2(7-p)}} \,   e^{2\,\beta_{p+1}\,\phi\,-\,2\,(7-p)\,B}  \nonumber\\
        &+& \frac{1}{2}\left[e^{-2B}\phi_r^2\, + \,  e^{-2D} \phi_z^2\right] \  + \  T \, e^{\gamma\,\phi} \ ; \label{eq:2_var_metric1}\\
rr&:& e^{-2B} \Big[(p+1)A_r^2- (B_r + F^+_r)^2 +(7-p)\left(B_r+\frac{1}{r}\right)^2+D_r^2\Big] +\frac{e^{-2B}}{r^2} \, \tilde{R}\nonumber \\
    &+&  e^{-2D}\Big[2 (B-D-F^-)_{zz} -(p+1) A_z^2 -(B_z-F_z^-)^2 - (7-p)B_z^2 + D_z^2\Big]\nonumber \\
        &=&   \frac{1}{2}\left[- \ e^{-2B}\phi_r^2\ + \  e^{-2D}  \phi_z^2\right] \  +  \frac{h_0^2}{2 \, r^{2(7-p)}} \,   e^{2\,\beta_{p+1}\,\phi\,-\,2\,(7-p)\,B} \nonumber \\
        &+& \frac{1}{2} \left[e^{-2B} {\cal B}_r^2 \ - \  e^{-2D}  {\cal B}_z^2 \right] \, e^{-\,2\,\beta_p\,\phi\,-\,2\,(p+1)\,A} \ + \  T \, e^{\gamma\,\phi} \ ; \label{eq:2_var_metric2} \\
rz &:& - \left(F^+ + F^-\right)_{rz} + \left(F^+ + F^-\right)_z  D_r\nonumber   \\
    &+&  \left(F^+ + F^-\right)_r B_z - 2(p+1) A_r \,A_z-2(7-p) \left(B_r+\frac{1}{r}\right)\, B_z  \nonumber \\
        &=&  \phi_r \,\phi_z\, - \,  e^{-\,2\,\beta_p\,\phi\,-\,2\,(p+1)\,A} \ {\cal B}_r \, {\cal B}_z   \ \label{eq:2_var_metric3} ;
\eea
\bea
zz &:&  \Big[ (p+1)A_z^2 + B_z^2 +(7-p) B_z^2 -(D_z + F_z^-)^2\Big]+ \frac{e^{2(D-B)}}{r^2} \, \tilde{R} \nonumber\\
    &+&  e^{2(D-B)}\Big[ 2\left(D-B-F^+\right)_{rr} -(p+1)A_r^2+ B_r^2 -(7-p)\left(B_r+\frac{1}{r}\right)^2-(D_r-F^+_r)^2\Big] \nonumber \\
                &=& \frac{1}{2}\ \left[\left(e^{D-B}\,\phi_r \right)^2 \,-\,  \phi_z^2\right]  - \frac{1}{2} \  \left[\left(e^{D-B}\,{\cal B}_r\right)^2 \,- \, {\cal B}_z^2
        \right] \, e^{-\,2\,\beta_p\,\phi\,-\,2\,(p+1)\,A} \nonumber \\
        &+&  e^{2D}\  \frac{h_0^2}{2 \, r^{2(7-p)}} \,   e^{2\,\beta_{p+1}\,\phi\,-\,2\,(7-p)\,B} \ + \ e^{2D} \  T \, e^{\gamma\,\phi} \ ; \label{eq:2_var_metric4} \\
mn&:& \frac{e^{-2B}}{r^2} \frac{5-p }{7-p} \ \tilde{R} \ + e^{-2B}\Big[2(\log{r}-F^+)_{rr}+2\left(B_r+\frac{1}{r}\right) F^+_r-(p+1)A_r^2 \nonumber \\
    &+& B_r^2 - (7-p)\left(B_r+\frac{1}{r}\right)^2 - D_r^2 - (F^+_r)^2 \Big]+ e^{-2D} \Big[2 (B-D-F^-)_{zz} \nonumber  \\
        &+& 2 F_z^- B_z -(p+1) A_z^2 -  B_z^2 - (7-p)  B_z^2+ D_z^2- ( F_z^-)^2  \Big] \nonumber \\
                    &=& \frac{1}{2}\left[e^{-2B}\phi_r^2\ + \  e^{-2D} \phi_z^2\right] \  -  \ \frac{h_0^2}{2 \, r^{2(7-p)}} \,   e^{2\,\beta_{p+1}\,\phi\,-\,2\,(7-p)\,B}   \nonumber \\
        &-& \frac{1}{2} \ \left[e^{-2B}  {\cal B}_r^2 \ + \  e^{-2D}  {\cal B}_z^2 \right] \, e^{-\,2\,\beta_p\,\phi\,-\,2\,(p+1)\,A} \ + \  T \, e^{\gamma\,\phi} \ , \label{eq:2_var_metric5}
\eea
where
\beq
\tilde{R}\ = \  (7-p)(6-p)
\eeq
is the curvature of the unit $(7-p)$-sphere.

\subsection{ \sc Exact Solutions}\label{sec:exact_solutions}

As we summarized in Section~\ref{sec:vacuum_solution}, the equations listed in the Section~\ref{sec:brane_equations} are clearly solved by the vacuum of~\cite{dm_vacuum}, which is characterized by ${\cal B}=0$ and by
\beq
A \ = \ B \ = \ D \ = \ \Omega(z) \ , \qquad \phi \ = \ \phi(z) \ ,
\eeq
where $\Omega(z)$ and $\phi(z)$ satisfy eqs.~\eqref{eq:background_eom}.

Interestingly, the uncharged branes of Section~\ref{sec:uncharged_branes} in the isotropic gauge of eq.~\eqref{eq:uncharged_isotropic}, with ${\cal D}=9$, vanishing values for $\phi_0$ and $\phi_1$ and line element $ds_9^2$, also yield exact solutions when dressed with $\Omega(z)$, so that the resulting backgrounds take the form
\beq \label{eq:9D_Ricci_flat}
ds^2 \ = \ e^{2\,\Omega(z)} \left(\, ds_9^2 \ + \ dz^2 \right) \ , \qquad \phi \ = \ \phi(z) \ .
\eeq
This is the case, since the nine--dimensional metric is related to the ten--dimensional one by a conformal transformation dictated by the vacuum solution,
\beq
g_{MN}^{(10)} \ = \ e^{2\Omega(z)} \ g_{MN}^{(9)} \ , \qquad g_{zz}^{(10)} \ = \ e^{2\Omega(z)} \ , \qquad g_{M z}^{(10)} \ = \ 0 \ ,
\eeq
with $M,N=0\ldots 8$, and where $g_{MN}^{(9)}$ is independent of $z$.
One can thus conclude that the scalar equation is satisfied, since $\phi$ only depends on $z$, and therefore $\Box\phi$ is not affected by the deformation of the nine--dimensional Minkowski vacuum.
Moreover, the ten-dimensional Ricci tensor can be expressed in terms of the nine-dimensional one as
\bea
{\cal R}^{(10)}_{MN} &=& {\cal R}^{(9)}_{MN} \ - \ \left[\Omega'' \ + \ 8 (\Omega')^2\right] g_{MN}^{(9)} \ , \nonumber \\
{\cal R}^{(10)}_{zz} &=& - \ 9 \Omega'' \ , \nonumber \\
{\cal R}^{(10)}_{Mz} &=& 0 \ ,
\eea
so that, in the absence of forms, the ten-dimensional equations
\bea
{\cal R}^{(10)}_{MN}  &=&    \frac{T}{8} \, e^{\gamma\phi} \, g^{(10)}_{MN}  \ , \nonumber \\
{\cal R}^{(10)}_{zz}  &=&  \frac{1}{2}\  (\phi')^2\ + \ \frac{T}{8}\, e^{\gamma\phi} \, g^{(10)}_{zz}  \ ,
\eea
reduce to
\bea
{\cal R}^{(9)}_{MN} &=&  \left[\Omega'' \ + \ 8 (\Omega')^2 + \frac{T}{8} \, e^{\gamma\phi+2\Omega}  \right]   g_{MN}^{(9)}  \ , \nonumber \\
0  &=& 9\Omega'' \ + \  \frac{1}{2}\  (\phi')^2\ + \ \frac{T}{8}\, e^{\gamma\phi+2\Omega} \,   \ .
\eea
Using eqs.~\eqref{eq:background_eom}, one is finally left with
\beq
{\cal R}^{(9)}_{MN}  \ = \ 0 \ .
\eeq

Brane isometries and the asymptotic behavior at large distances from their cores thus imply that the Ricci-flat nine-dimensional tadpole-free uncharged branes solutions of Section~\ref{sec:uncharged_branes}, when dressed with $\Omega$ as in eq.~\eqref{eq:9D_Ricci_flat}, capture the exact profiles of ten-dimensional uncharged branes in the presence of tadpole potentials.
Explicitly, the complete ten-dimensional solutions are then
\bea \label{eq:uncharged_branes_DM}
A &=& \Omega(z) \ \mp \ \sqrt{\frac{7-p}{7(p+1)}} \log\left[\frac{1+v(r)}{1-v(r)}\right] \ ,  \nonumber\\
B &=& \Omega(z) \ \pm \ \sqrt{\frac{(7-p)(p+1)}{7 (6-p)^2}}  \log\left[\frac{1+v(r)}{1-v(r)}\right] \ + \ \frac{1}{6-p} \log\left[1-v^2(r)\right] \ , \nonumber \\
D &=& \Omega(z) \ ,  \qquad \phi  \ = \  \phi(z) \ ,
\eea
where
\beq
v(r)\ = \ \left(\frac{ r_0 }{r}\right)^{6-p} \  .
\eeq
Here $r_0$ is a free positive parameter, and the upper (lower) signs correspond to positive (negative) values for the brane tension
\beq
{\cal T}_p \ = \ \pm \, 14 \, \sqrt{\frac{7-p}{7(p+1)}} \frac{\Omega_{7-p}}{\kappa_9^2} \, r_0^{6-p} \ ,
\eeq
where $\kappa_9^2$ is the effective nine-dimensional gravitational constant, which is linked to the ten-dimensional one according to
\beq
\frac{1}{\kappa_9^2} \ = \ \frac{1}{\kappa_{10}^2} \ \int_0^{z_m} dz \, e^{8\Omega}\ .
\eeq
Note that in the two cases of interest, where $\Omega$ has the limiting behaviors as in eqs.~\eqref{omegaphi_orientifold} and~\eqref{omegaphi_heterotic}, the integral on the right hand side is finite.
These solutions actually describe ten--dimensional $(p+1)$-branes, one of whose directions extends along the internal $z$-interval.

More general gauge choices could lead to analytic expressions for other classes of exact solutions. In fact, this occurs, in the presence of tadpole potentials, for the vacua of~\cite{dm_vacuum}, which afford analytic solutions in a special gauge that eliminates the exponential potentials, and for the general charged branes in the absence of tadpoles described in~\cite{mrs24_1}, for which similar considerations apply in the harmonic gauge.
The more general ansatz
\bea
ds^2 &=& e^{2A(\zeta,u)}dx_{p+1}^2 \, + \, e^{2B(\zeta,u)} \Big(d u \,+\, E(\zeta,u) \,d\zeta \Big)^2 \, + \, e^{2D(\zeta,u)} d\zeta^2 \,+\, e^{2C(\zeta,u)} d\,\Omega^2_{7-p}  \ , \nonumber \\
\phi &=& \phi(\zeta,u) \ ,
\eea
where now $(z,r)$ are arbitrary functions of $(\zeta,u)$, can prove instrumental to this effect. It rests on two additional functions, $E$ and $C$, with respect to the ansatz in eqs.~\eqref{eq:ansatz_2variables1}, which brings along a two--parameter gauge symmetry.
The complete equations of motion, to which we plan to return in the future, can be found in Appendix~\ref{app:brane_details}, but in the following we shall content ourselves with exploring the linearized approximation, which should capture the general behavior at large distance from brane cores.

\section[Solutions of the Linearized System]{\sc The Linearized System and Large--Distance Behavior}
\label{sec:linearized_analysis}

Since, as we have stressed, generic brane solutions are to approach the background as $r \to \infty$, one can examine the asymptotics expanding the different functions around the
vacuum solution according to
\bea
A &=& \Omega(z) \ + \ a(z,r) \ , \qquad B \ = \ \Omega(z)\ + \ b(z,r) \ ,\nonumber \\
D&=&\Omega(z)\ + \ d(z,r) \ , \qquad \phi \ = \ \phi(z) \ + \ \varphi(z,r)
\eea
and linearizing the preceding equations in the perturbations $a$, $b$, $d$, $\varphi$, which all vanish as $r\to\infty$.

\subsection{\sc The Linearized Equations}

The linearized $r z$ equation in~\eqref{eq:2_var_metric3} reads
\beq
-  \ 2\left[(p+1) \, a \ + \ (7-p)\,b\right]_{rz}\  + \  16\, \Omega_z \, d_r = \varphi_r \,\phi_z \ .
\eeq
Taking the boundary conditions as $r\to\infty$ into account, one can integrate it so that it determines $\varphi$ in terms of the other fields, according to
\beq \label{eq:dilaton_two_variables}
\varphi \ = \ - \ \frac{2}{ \phi_z} \Big[(p+1) \, a_z \  + \  (7-p)\, b_z  \ - \ 8\, \Omega_z\,  d\Big] \ .
\eeq
In the following it will be convenient to define the combinations
\bea
\xi & = & \frac{(p+1)\,a \ + \ (6-p)\, b\ + \ d}{2} \ , \nonumber\\
\rho & = & b \ - \ a \ , \nonumber\\
\chi & = & (p+1)\, a \  + \  (7-p)\, b \ .
\eea
Using the background equations of motion~\eqref{eq:background_eom}, and expressing $\varphi$ in terms of the other perturbations via eq.~\eqref{eq:dilaton_two_variables}, the linearized $(\mu\nu)$ Einstein equations~\eqref{eq:2_var_metric1} and the difference between them and the $(rr)$ equation~\eqref{eq:2_var_metric2} can be turned to
\bea
(\mu\nu)&:&  (2\xi+\rho)_{rr} \ + \ \frac{7-p}{r}\left(2\xi+\rho\right)_r \ + \ \rho_{zz} +8  \Omega_z  \rho_z \ =\  0 \ ,\\
(\mu\nu) - (rr)&:&   (2\xi+\rho)_{rr}\ +\ \frac{7-p}{r}\rho_r+\rho_{zz} \ + \ 8\Omega_z\rho_z  \ =\  0 \ . \label{eq:mu_r_eq}
\eea
Combining these two equations, one can then conclude that
\beq
\xi_r = 0 \ ,
\eeq
so that $\xi$ must vanish altogether, in view of the boundary conditions as $r\to\infty$, and consequently
\beq
d \ = \ - \ (p+1)\,a \ - \ (6-p) \,b \ .
\eeq

When $\xi=0$, the $(mn)$ and $(\mu\nu)$ Einstein equations~\eqref{eq:2_var_metric5}  and~\eqref{eq:2_var_metric1} coincide, the combination in~\eqref{eq:dilaton_two_variables} automatically satisfies the dilaton equation, and one is thus left with
\bea \label{eq:rho_chi}
(\mu\nu)&:&  \rho_{rr} \ +  \ \rho_{zz} \  +\ \frac{7-p}{r}\,\rho_r \ + \ 8  \, \Omega_z \, \rho_z \ = \  0 \ , \nonumber\\
(zz) &:& \chi_{rr} \ + \ \chi_{zz} \ + \ \frac{7-p}{r} \, \chi_r  \ + \ \frac{2}{ \phi_z} \left(12 \, \Omega_z \, \phi_z \ - \ \gamma \, T  \, e^{2\Omega+\gamma\phi}\right)\chi_z  \nonumber \\
& & - \ \frac{7}{4} \left(1\,+\,\frac{8\, \gamma \,\Omega_z}{ \phi_z}\right) T \,e^{2\Omega+\gamma\phi} \ \chi\ = \  - \ \frac{p+1}{4} \left(1\,+\,\frac{8\, \gamma \,\Omega_z}{\phi_z}\right) T \, e^{2\Omega+\gamma\phi} \ \rho \ .
\eea

Finally, the equations of motion of the form field are completely decoupled from the others and read
\beq \label{eq:linearized_form}
{\cal B}_{zz}- \Big[2\beta_p\phi \ + \  2(p-3)\Omega\Big]_z {\cal B}_z  \ = \ - \ {\cal B}_{rr} -  \frac{7-p}{r} {\cal B}_{r} \ .
\eeq

\subsection{\sc The Linearized Dilaton--Gravity Sector}

Let us begin by considering the linearized equations for the gravity sector.

\subsubsection{\sc The $\rho$ equation}

The first of eqs.~\eqref{eq:rho_chi} determines $\rho$, which then enters the second as a source.
One can separate variables in the equation for $\rho$ letting
\beq
\rho(z,r)  \ = \ \rho_1 (r) \, \rho_2(z) \ ,
\eeq
which leads to
\beq
(\rho_1)_{rr}\ + \ \frac{7-p}{r}(\rho_1)_r \ = \ m^2 \ \rho_1 \ , \qquad
(\rho_2)_{zz}\ + \ 8\,\Omega_z(\rho_2)_z \ = \  - \ m^2 \ \rho_2 \ ,  \label{eqs_rho}
\eeq
where $m^2$ is a constant.

The second of eqs.~\eqref{eqs_rho} can now be turned into the Schr\"odinger form
\beq
- \ \psi_{zz} + \left[4 \, \Omega_{zz}\ + \ 16\, \Omega_z^2 \ - \  m^2\right]\psi \ = \ 0 \ , \label{schrod_psi}
\eeq
letting
\beq \label{eq:psirho2}
\psi \ = \  e^{4 \,\Omega} \ \rho_2 \ .
\eeq
Eq.~\eqref{schrod_psi} determines the possible values of $m^2$, and can be cast in the formally positive form
\beq
{\cal A}^\dagger {\cal A} \ \psi \ = \ m^2\, \psi \ , \label{ham_psi}
\eeq
where
\beq
{\cal A} \ = \ - \ \frac{d}{dz} \ + \ 4\, \Omega_z \ .
\eeq

All the Schr\"odinger potentials that will emerge in this section, and in particular the potential in eq.~\eqref{schrod_psi}, have double--pole singularities in the conformal variable $z$ near the endpoints of the internal interval, where they behave as
\beq \label{eq:pontential_endpoints}
V(z) \ \sim \ \frac{\mu^2 \,-\,\frac{1}{4}}{z^2} \ , \qquad V(z) \ \sim \ \frac{\tilde{\mu}^2 \,-\,\frac{1}{4}}{\left(z_m-z\right)^2} \ .
\eeq
Singular potentials of this type were discussed at length in~\cite{ms23_1,ms23_2}, and can be conveniently approximated by hypergeometric functions. Near the endpoints, the corresponding Schr\"odinger wavefunctions $\psi$ behave as
\bea
\psi &\sim& C_1 \ \left(\frac{z}{z_m}\right)^{\frac{1}{2}+\mu} \ + \ C_2 \left(\frac{z}{z_m}\right)^{\frac{1}{2}-\mu} \ , \nonumber \\
\psi &\sim& C_3 \left(1-\frac{z}{z_m}\right)^{\frac{1}{2}+\tilde{\mu}} \ + \ C_4 \left(1-\frac{z}{z_m}\right)^{\frac{1}{2}-\tilde{\mu}} \ . \label{psi_limit}
\eea
These results capture almost all cases of interest, but for $\mu=0$ the limiting behavior near $z=0$ include logarithms, and is described by
\beq
\psi \ \sim \ C_1 \ \left(\frac{z}{z_m}\right)^{\frac{1}{2}} \ + \ C_2 \left(\frac{z}{z_m}\right)^{\frac{1}{2}} \log \left(\frac{z}{z_m}\right) \ .
\eeq
Similar results hold near $z_m$ for $\tilde{\mu}=0$.
When $\mu$ (or $\tilde{\mu}$) are larger than one, $C_2$ (or $C_4$) must vanish for normalizable modes. On the other hand, when $0<\left(\mu,\tilde{\mu}\right)<1$ the ratios $\frac{C_2}{C_1}$ and $\frac{C_4}{C_3}$ parametrize the different self--adjoint boundary conditions. Generic choices, however, can violate the formal positivity of eq.~\eqref{ham_psi}, as explained in detail in~\cite{ms23_1}.

Once the allowed values of $m^2$ are known, the first of eqs.~\eqref{eqs_rho} determines the limiting behavior of the perturbations as $r\to\infty$.
When $m=0$, after taking the boundary conditions into account, the solution for $\rho_1$ that decays at infinity is
\beq
\rho_1 \ = \ \frac{\rho_0}{r^{6-p}} \ .
\eeq

Negative values of $m^2$ signal instabilities, which manifest themselves via their generic exponential growth in time, but for the vacuum solutions that we consider it is always possible to find boundary conditions that only allow stable perturbations.
On the other hand, when $m^2$ is positive, the solutions decay exponentially for large values of $r$, and the exact solution for $\rho_1$ that decays at infinity is
\beq
\rho_1  \ = \ \rho_0 \, r^{-\,\frac{6-p}{2}} K_{\frac{6-p}{2}} \left( m r \right) \ , \label{modified_bessel}
\eeq
where $K$ is a modified Bessel function that decreases exponentially as $r \to \infty$ and $\rho_0$ is a constant. Indeed, for large values of $r$
\beq
\rho \ \sim \ r^{-\,\frac{7-p}{2}} \ e^{-\, m r} e^{-\,4\,\Omega(z)}\ \psi_m(z) \ ,
\eeq
with $\psi_m$ the proper Schr\"odinger wavefunction with eigenvalue $m$.

Turning to the cases of direct interest here, the orientifold background behaves near the endpoints as in eqs.~\eqref{omegaphi_orientifold}, and consequently in this case
\beq
\mu \ = \ 0 \ , \qquad \tilde{\mu} \ = \ 0 \ ,
\eeq
so that independent boundary conditions are allowed, in principle, at the two ends. However, the analysis in~\cite{ms23_1} showed that a stable spectrum forbids logarithmic limiting behaviors, and thus demands that the Schr\"odinger wavefunction $\psi$ approaches $\left(\frac{z}{z_m}\right)^\frac{1}{2}$ and $\left(1 - \frac{z}{z_m}\right)^\frac{1}{2}$ at the two ends. In fact, eq.~\eqref{ham_psi} yields directly a normalizable zero mode of this type,
\beq
\psi \ = \  e^{4 \,\Omega}\, \psi_0 \ ,
\eeq
with $\phi_0 $ constant. This corresponds to a constant $\rho_2$, and is accompanied by a power--like behavior for $\rho_1 \sim r^{-\left(6-p\right)}$, so that the zero--mode solution for $\rho$ is simply
\beq
\rho \ = \ \frac{\rho_0}{r^{6-p}} \ ,
\eeq
where $\rho_0$ is a constant. On the other hand, massive solutions decay exponentially for large values of $r$, but still approach constant values at the two ends of the internal interval, in compliance with the preceding discussion about boundary conditions.
\begin{figure}[ht]
\centering
\includegraphics[width=75mm]{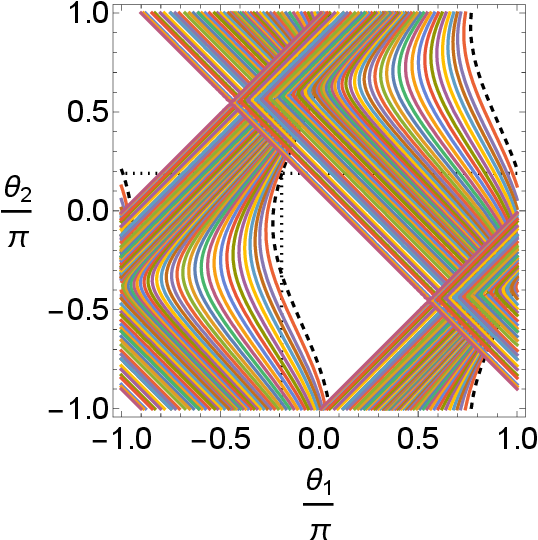}
\caption{\small The shaded regions in the figure identify the boundary conditions, which are parametrized by $(\theta_1,\theta_2)$, where $\frac{C_1}{C_2}=\tan\left(\frac{\theta_1-\theta_2}{2}\right)$ and $\frac{C_3}{C_4}=\tan\left(\frac{\theta_1+\theta_2}{2}\right)$, which lead to instabilities for this sector, as discussed in~\cite{ms23_2}. The dashed lines in the white regions correspond to boundary conditions with zero modes but no instabilities.}
\label{fig:grav_instabilities_130}
\end{figure}
For the heterotic SO(16) $\times$ SO(16) model, the limiting behavior is given in eqs.~\eqref{omegaphi_heterotic}, and consequently the potential behaves as in eq.~\eqref{eq:pontential_endpoints} with
\beq
\mu \ = \ \frac{1}{3} \ , \qquad \tilde{\mu} \ = \ 0 \ .
\eeq
Therefore, the behavior of the wavefunction near the two endpoints is
\bea
\psi & \sim & C_1 \left(\frac{z}{z_m}\right)^{\frac{5}{6}}  \ + \ C_2 \left(\frac{z}{z_m}\right)^{\frac{1}{6}} \ ,  \nonumber \\
\psi & \sim & C_3 \left(1-\frac{z}{z_m}\right)^{\frac{1}{2}} \ + \ C_4 \left(1-\frac{z}{z_m}\right)^{\frac{1}{2}} \log\left(1-\frac{z}{z_m}\right) \ , \label{psi_beh}
\eea
and the self--adjoint boundary conditions are parametrized by the two ratios $\frac{C_1}{C_2}$ and $\frac{C_4}{C_3}$.
The corresponding behavior for $\rho_2$ is
\bea \label{eq:rho_het}
\rho_2 & \sim & C_1 \left(\frac{z}{z_m}\right)^{\frac{2}{3}}  \ + \ C_2  \ ,  \nonumber \\
\rho_2 & \sim & C_3 \ + \ C_4  \log\left(1-\frac{z}{z_m}\right) \ ,
\eea
so the zero mode of eq.~\eqref{eq:psirho2} corresponds to equal values of $C_2$ and $C_3$, with $C_4=0$.
There is an additional normalizable zero mode in this case, which can be deduced relying on the Wronskian method. The corresponding general result for $\rho_2$ reads
\beq
A_1  \ + \ A_2  \int_{z_0}^z d z' e^{\,-\,8\Omega(z')} \ . \label{psi_2_massless}
\eeq
The limiting behavior at the two ends is as in eq.~\eqref{eq:rho_het}, with the ratio between $C_4$ and $C_3$ determined by ratio between $A_1$ and $A_2$, which are linked to $C_1$ and $C_2$.
A stability analysis along the lines of~\cite{ms23_1}, but relying on hypergeometric functions as in~\cite{ms23_2}, yields results that are very similar to those presented there for the four--dimensional graviton--dilaton sector of that case, which shares the same values of $\mu$ and $\tilde{\mu}$. The resulting stability plot is shown in fig.~\ref{fig:grav_instabilities_130}, and is almost identical to that displayed in fig.~5 of~\cite{ms23_2}, and indicates that there are $z$--dependent massless modes for other choices of boundary conditions.
Fig.~\ref{fig:grav_instabilities_130} also indicates that stable boundary conditions lie in the region
\beq
- \ 1.14 \ < \ \frac{C_1}{C_2} \ < \ 0 \ , \label{stability_so1616}
\eeq
which translates into a corresponding stability region for $A_1$ and $A_2$.
In general, the zero mode solution for $\rho$ at large values of $r$ becomes
\beq
\frac{A_1}{r^{6-p}}  \ + \ \frac{A_2}{r^{6-p}}  \int_{z_0}^z d z' e^{\,-\,8\Omega(z')}  \ .
\eeq

\subsubsection{\sc The $\chi$ equation}

The second equation in~\eqref{eq:rho_chi} can be solved for $\chi$ adding to the general solution of the corresponding homogeneous equation a special solution induced by $\rho$, which behaves as a source. Therefore
\beq
\chi \ = \ \chi_0 \ + \  \chi_s \ ,
\eeq
where $\chi_0$ solves the homogeneous equation
\bea
& &(\chi_0)_{rr}\ + \ (\chi_0)_{zz}\ + \ \frac{7-p}{r} \, (\chi_0)_r  \ + \ \frac{2}{\phi_z} \left(12 \Omega_z \phi_z - \gamma T e^{2\Omega+\gamma\phi}\right)\,(\chi_0)_z  \nonumber \\
& & - \ \frac{7}{4} \left(1+\frac{8 \,\gamma \,\Omega_z}{\phi_z}\right)   T \, e^{2\Omega+\gamma\phi} \, \chi_0\ = \ 0 \ ,
\eea
while $\chi_s$ is a special solution of the complete inhomogeneous one. The latter can be determined in this case starting from a factorized form,
\beq
\chi_s(z,r) \ = \ \rho_1(r) \chi_{s,2}(z) \ ,
\eeq
with the same $r$-dependence as the source. Inserting this expression in eq.~\eqref{eq:rho_chi} leads to
\bea
&&\!\!\!\!\! \left(\chi_{s,2}\right)_{zz}  \ + \ \frac{2}{ \phi_z} \left(12 \, \Omega_z \, \phi_z \ - \ \gamma \, T  \, e^{2\Omega+\gamma\phi}\right)\left(\chi_{s,2}\right)_z  \nonumber \\
&+&\!\!\!\!\!\left[ m^2 \ - \  \frac{7}{4} \left(1\,+\,\frac{8\, \gamma \,\Omega_z}{ \phi_z}\right) T \,e^{2\Omega+\gamma\phi} \right] \chi_{s,2}\ = \  - \ \frac{p+1}{4} \left(1\,+\,\frac{8\, \gamma \,\Omega_z}{\phi_z}\right) T \, e^{2\Omega+\gamma\phi} \rho_2 \, .
\eea
The special solution can be expressed in terms of the Green function $G(z,z')$, defined according to
\bea
&&\!\!\!\!\! G_{zz}(z,z')  \ + \ \frac{2}{ \phi_z} \left(12 \, \Omega_z \, \phi_z \ - \ \gamma \, T  \, e^{2\Omega+\gamma\phi}\right)G_z(z,z')  \nonumber \\
&+&\!\!\!\!\!\left[ m^2 \ - \  \frac{7}{4} \left(1\,+\,\frac{8\, \gamma \,\Omega_z}{ \phi_z}\right) T \,e^{2\Omega+\gamma\phi} \right] G(z,z')\ = \ \delta(z-z') \, ,
\eea
which leads to an implicit expression for $\chi_{s,2}$
\beq
\chi_{s,2}(z) \ = \ - \  \int dz' \, G(z,z') \, \frac{p+1}{4} \left(1\,+\,\frac{8\, \gamma \,\Omega_z(z')}{\phi_z(z')}\right) T \, e^{2\Omega(z')+\gamma\phi(z')} \, \rho_2 (z') \ .
\eeq
For the constant zero mode of $\rho_2$, which is allowed in both orientifold and heterotic models, the special solution $\chi_{s,2}$ can be determined explicitly, and is constant, so that
\beq
\chi_s \ = \ \frac{p+1}{7} \frac{\rho_0}{r^{6-p}} \ .
\eeq
This result translates into
\beq \label{eq:linear_uncharged}
a \ = \ - \frac{6-p}{7}\frac{\rho_0}{r^{6-p}} \ , \qquad b \ = \ \frac{p+1}{7}\frac{\rho_0}{r^{6-p}} \ , \qquad d \ = \ 0 \ , \qquad \varphi \ = \ 0 \ ,
\eeq
which capture precisely the large--$r$ limit of the uncharged branes of eqs.~\eqref{eq:uncharged_branes_DM} in the nine--dimensional Minkowski background with
\beq
\rho_0 \ = \ \pm \, \frac{2}{6-p}\,\sqrt{\frac{7(7-p)}{(p+1)}}\,r_0^{6-p} \ .
\eeq
The brane tension is determined by $\rho_0$ according to
\beq
{\cal T}_p  \ = \ (6-p)\,\frac{\Omega_{7-p}}{\kappa_9^2} \,\rho_0 \ .
\eeq
On the other hand, massive modes of the orientifold vacua yield special $z$-dependent solutions.
The behavior at the two ends is obtained using that $C_4=C_2=0$, so that the special solution behaves as $\left(\frac{z}{z_m}\right)^\frac{9}{2}$ and $\left(1 - \frac{z}{z_m}\right)$ at the two ends. In the SO(16) $\times$ SO(16) case, where in general eq.~\eqref{psi_2_massless} determines the zero mode, the special solution $\chi_s$ depends on $z$. At the two ends it approaches a constant, unless $C_3=0$ ($C_1$ cannot vanish, for stable boundary conditions, in view of eq.~\eqref{stability_so1616}), in which case it vanishes at the right end.

Turning now to the homogeneous equation for $\chi(z,r)$, one can separate variables letting
\beq
\chi_0  \ = \ \chi_1(r) \,\chi_2(z) \ .
\eeq
The equation for $\chi_1(r)$ becomes analogous to the first of eqs.~\eqref{eqs_rho},
\bea
(\chi_1)_{rr} \ + \ \frac{7-p}{r}\,(\chi_1)_r \ = \  M^2 \chi_1 \ ,
\eea
whose solution is again a Bessel function for $M \neq 0$ or a power for $M=0$. The equation for $\chi_2(z)$ becomes
\beq
(\chi_2)_{zz} \ +\  \left[8 \Omega_z - 2 \frac{\phi_{zz}}{\phi_z} \right] \, (\chi_2)_z \ + \  \left[M^2 \ - \ \frac{7}{4} \left(1+\frac{8 \,\gamma \,\Omega_z}{\phi_z}\right)  T \, e^{2\Omega+\gamma\phi}\right]\chi_2   \ = \  0\ .
\eeq
Letting
\beq
\chi_2 \ =\  \phi_z \ e^{-4 \Omega} \  \Psi \ , \label{chi2psi}
\eeq
the resulting equation for $\Psi$ is of the Schr\"odinger form, with a potential
\beq
V \ = \  \alpha_z \ + \ \alpha^2 \ + \ b_0 \ ,
\eeq
where
\beq
\alpha \ = \ 4\,\Omega_z  - \frac{\phi_{zz}}{\phi_z} \ , \qquad b_0 = \frac{7}{4} \left(1+\frac{8 \,\gamma \,\Omega_z}{\phi_z}\right) T \, e^{2\Omega+\gamma\phi} \ .
\eeq
The Schr\"odinger equation can be cast in the equivalent form
\beq\label{eq:chi_0_hamiltonian}
\left[ \left(\frac{d}{d z} + \alpha\right)\left( - \ \frac{d}{d z} + \alpha\right)\ + \ b_0 \right] \Psi \ = \ M^2 \, \Psi \ ,
\eeq
and in both cases relevant for ten-dimensional non-supersymmetric strings the exact solutions of~\cite{dm_vacuum} with $\gamma=\frac{3}{2}$ or $\gamma=\frac{5}{2}$ imply that $b_0$ is positive. Consequently, eq.~\eqref{eq:chi_0_hamiltonian} describes a formally positive Hamiltonian since it is the sum of two positive operators.

For the orientifolds, the potential has double poles at the ends of the interval, as in eq.~\eqref{eq:pontential_endpoints}, with
\beq
\mu \ = \ 1 \ , \qquad \tilde{\mu} \ = \ 1 \ ,
\eeq
so that there is a unique self--adjoint boundary condition for $\Psi$ at both ends, such that
\beq
\Psi \ \sim \ \left(\frac{z}{z_m}\right)^\frac{3}{2} \ , \qquad \Psi \ \sim \ \left(1 \ - \ \frac{z}{z_m}\right)^\frac{3}{2} \ . \label{Psi_orientifold}
\eeq
The corresponding solution of the homogeneous equation of $\chi_2$ for the orientifolds thus approaches a constant at both ends of the internal interval, $M^2>0$ and thus $\chi_1$ decays exponentially.
Therefore, for large values of $r$ one is left with $\chi_s$ only, and the solution approaches the same behavior as the uncharged branes in the lower-dimensional Minkowski background of eqs.~\eqref{eq:linear_uncharged}.
In general, this asymptotic behavior may be common to several non-linear complete solutions, possibly $z$-dependent.
\begin{figure}[ht]
\centering
\includegraphics[width=80mm]{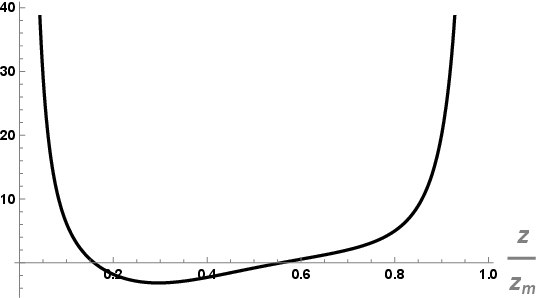}
\caption{\small The potential for scalar perturbations in the heterotic SO(16) $\times$ SO(16) model.}
\label{fig:het_scalar_potential}
\end{figure}
For the heterotic case, the potential is characterized by
\beq
\mu \ = \ \frac{2}{3} \ , \qquad \tilde{\mu} \ = \ 1 \ ,
\eeq
so that there is a unique limiting behavior at the right end, as in eq.~\eqref{Psi_orientifold}, while close to the left end $\Psi$ behaves as
\beq
\Psi \ \sim \ C_1 \left(\frac{z}{z_m}\right)^\frac{7}{6} \ + \ C_2 \left(\frac{z}{z_m}\right)^{-\,\frac{1}{6}} \ .
\eeq
The potential is shown in fig.~\ref{fig:het_scalar_potential}, and is almost identical to the one discussed in fig.~19 of~\cite{ms23_2}. Also in this case, the sign of $M^2$ depends on the choice of boundary conditions, and the correspondence with the results presented there indicates that when $\frac{C_2}{C_1}$ is close to zero the spectrum is free of instabilities. In this case, eq.~\eqref{chi2psi} shows that $\chi_2(z)$ is essentially constant in the internal interval. The ensuing discussion then resembles what we said on the orientifold case.

\subsection{\sc The Linearized {\cal B}-Tensor Spectrum}

Let us now consider the equations of the form fields, eqs.~\eqref{eq:linearized_form}.
One can separate variables for ${\cal B}$, letting
\beq
{\cal B}(z,r) = {\cal B}_1(r) \, {\cal B}_2(z) \ ,
\eeq
which leads to the two ordinary differential equations
\bea
{{\cal B}_1}_{rr} &+&\frac{7-p}{r}\, {{\cal B}_1}_r\ =\ m^2 {{\cal B}_1} \ , \nonumber\\
{{\cal B}_2}_{zz}&+& \Big[-2\,\beta_p\,\phi_z \ + \  2(3-p)\,\Omega_z\Big]{{\cal B}_2}_z  \ = \ -  \ m^2 {{\cal B}_2} \ . \label{tensor_separate}
\eea
From what we saw when we discussed the first of eqs.~\eqref{eqs_rho}, for $m \neq 0$ the relevant solution for ${\cal B}_1$ is related to the modified Bessel function given in eq.~\eqref{modified_bessel}. However, for charged branes $p$ is odd, and there is a simplification, since the Bessel functions of half--odd--integer order are combinations of exponentials and powers.

The second equation in~\eqref{tensor_separate} is an eigenvalue equation for $m^2$, and can be turned into the Schr\"odinger form
\beq
-\ \psi_{zz} \ + \ V\psi \ = \ m^2 \, \psi \ ,
\eeq
by the redefinition
\beq
{{\cal B}_2} \ =  \ \psi  \  e^{\beta_p\phi+(p-3)\Omega} \ ,
\eeq
so that the potential is
\beq
V\ =\ - \
\Big[\beta_p\,\phi_{zz}+ (p-3)\Omega_{zz}\Big] \ + \ \Big[\beta_p\,\phi_{z}+ (p-3)\Omega_{z}\Big]^2 \ .
\eeq
The Hamiltonian has the manifestly positive ${\cal A}^\dagger {\cal A}$ form, with
\beq
{\cal A} \ = \ \frac{d}{dz} \ + \ \left[\beta_p \, \phi_z+(p-3)\Omega_z\right] \ .
\eeq

We can now examine in detail the behavior of tensor modes, treating separately the orientifolds and the heterotic SO(16)$\times$SO(16) model.

\subsubsection{\sc Orientifold Models}

For the orientifold models
\beq
\beta_p \ = \ \frac{p-3}{4} \ , \label{betap_or}
\eeq
and the limiting behaviors~\eqref{omegaphi_orientifold} for the background imply that, close to the two ends, the potential behaves as in eq.~\eqref{eq:pontential_endpoints}, with
\beq
\mu \ = \ \frac{\left|p \,-\,2\right|}{2} \ , \qquad \tilde{\mu} \ = \ \frac{\left|p \,-\,5\right|}{4} \ .
\eeq
Therefore, referring to eq.~\eqref{psi_limit} and the following discussion, normalizability demands that $C_4=0$ for $p=1$ and $C_2=0$ for $p=5$, while all the $C_i$'s are allowed for $p=3$.

We can now examine, in all these cases, the massless modes. To begin with, the solution of ${\cal A}\psi = 0$ reads
\beq
\psi  \  = \  e^{-\,\beta_p\phi- (p-3)\Omega} \ . \label{eq:form_psi_zeromode}
\eeq
Making use of the limiting behavior in  eqs.~\eqref{omegaphi_orientifold} and taking the values of $\beta_p$ in~\eqref{betap_or} into account, one finds that, close to the two ends
\beq
\psi\sim z^{-\frac{p-3}{2}} \ , \qquad \psi \sim (z_m-z)^{ \, \frac{p-3}{4}} \ ,
\eeq
so that the solution~\eqref{eq:form_psi_zeromode} is normalizable only for $p=3$. However, for $p=1$ one can define a normalizable Schr\"odinger zero-mode by the Wronskian method, as
\beq
\tilde{\psi} (z)\ = \ \psi(z) \ \int_{z_m}^z \frac{dz'}{\left[\psi(z')\right]^2} \ , \label{wrosnk_zm}
\eeq
which behaves as $(z_m-z)^{\frac{3}{2}}$ near the right endpoint and approaches a constant at the origin. On the other hand, for $p=5$ one can again define a normalizable Schr\"odinger zero-mode by the Wronskian method, as
\beq
\tilde{\psi} (z)\ = \ \psi(z) \ \int_{0}^z \frac{dz'}{\left[\psi(z')\right]^2} \ ,
\eeq
which behaves as $z^{2}$ at the origin and logarithmically at the other end. Summarizing, there are normalizable zero modes for all values of $p$ of interest for the two orientifold models, with a power--like behavior proportional to $r^{p-6}$ at large distances.

For the USp(32) model, the relevant values of $p$ are 1 and 5, and in both cases ${\cal B}$ depends on $z$. In detail:
\bea
p=1 : \qquad \quad {\cal B} &=&  \frac{Q_1}{r^{5}} \ \int_z^{z_m}  {dz'}\ e^{-\,4\,\Omega(z')\,-\,\phi(z')} \ , \nonumber \\
p=5 : \qquad \quad {\cal B} &=&  \frac{Q_5}{r} \ \int_{0}^z  {dz'}\ e^{4\,\Omega(z')\,+\,\phi(z')} \ ,
\eea
which determine the corresponding field strength profiles according to eqs.~\eqref{eq:H_and_H_tilde} and~\eqref{eq:two_variable_ansatz_p_brane},
\bea
{\cal H}_3 &=& -\frac{Q_1}{r^5} \left[\frac{5 dr}{r}\int_z^{z_m}  {dz'}\ e^{-\,4\,\Omega(z')\,-\,\phi(z')} \ + \ dz\  e^{-\,4\Omega(z)-\phi(z)}\right]\wedge dx^0\wedge dx^1 \ , \nonumber\\
{\cal H}_7 &=&\frac{Q_5}{r}\left[-\frac{ dr}{r}\int_{0}^z  {dz'}\ e^{4\,\Omega(z')\,+\,\phi(z')} \ + \  dz\  e^{4\Omega(z)+\phi(z)}\right]\wedge dx^0\wedge\ldots\wedge dx^5 \ .
\eea
For the U(32) model, there is the additional option $p=3$. In this case eq.~\eqref{tensor_separate} becomes free, because of conformal symmetry, and the solutions for ${\cal B}_2$ are elementary. For example, Neumann boundary conditions at both ends of the interval grant the existence of a zero mode, together with a whole spectrum of massive modes with
\bea \label{eq:tensor_B1B2}
{\cal B}_{1,\ell}(r) &=& \frac{1\,+\, m_\ell \, r}{r^3} \ e^{-\,m_\ell \, r} \ ,  \, \qquad {\cal B}_{2,m_\ell}(z) \ = \ C_\ell \ \cos\left(m_\ell \, z\right) \ , \nonumber \\
 m_\ell^2 &=& \left(\frac{\ell \,\pi}{z_m}\right)^2 \qquad \ell=0,1,\dots \ ,
\eea
so that the mass gap is $\frac{\pi}{z_m}$.
Other choices of boundary conditions remove the zero mode. For example, Dirichlet boundary conditions yield sin functions with the same spectrum, up to the removal of the $\ell=0$ mode.

If the internal manifold were a circle of length $2 {z_m}$, the form-field solution would involve the harmonic function $\frac{1}{\left(r^2+z^2\right)^2}$ extended periodically along it, and then the Poisson summation formula would yield the decomposition
\beq
\sum_{k \in Z} \frac{1}{\left[ r^2 \ + \ \left(z - 2 k{z_m} \right)^2\right]^2} \ = \ \frac{\pi}{4\,z_m\,r^3}\left[1 \,+\, 2 \sum_{\ell=1}^\infty e^{-\,\frac{\pi \, \ell \, r}{z_m}} \left(1 + \frac{\pi \, \ell \, r}{z_m}\right) \cos\left( \frac{\pi \ell \, z}{z_m}\right) \right] \ , \label{poisson}
\eeq
which also applies to the Neumann-Neumann case. Other boundary conditions result in similar, if more complicated, decompositions.
If one demands that the dependence on $r$ and $z$ become isotropic for large values of $z_m$, the leading behavior of the $C_\ell \sim \frac{\pi}{2\,z_m}$ for large values of $\ell$ should hold, independently of the choice of boundary conditions, if $z$ lies in the interior of the interval. An infinite number of modes will contribute to the actual ${\cal B}$ profiles for charged brane solutions.

Returning to the zero mode, one can define the large--$r$ behavior of the $p=3$ form field strength starting from
\beq \label{eq:form_2_variable_10D}
{\cal H}_{5} \ = \ \frac{Q_3}{r^{4}} \ dx^0 \wedge \ldots \wedge dx^3 \wedge dr \ ,
\eeq
but the self-duality condition requires that it be completed according to
\beq
{\cal H}_5^\prime \ = \ \frac{1}{\sqrt{2}} \left({\cal H}_5 \ + \ \star {\cal H}_5 \right)  \ ,
\eeq
where ${\cal H}_5$ is given in eq.~\eqref{eq:form_2_variable_10D}.
Explicitly, one thus finds that
\beq
{\cal H}_5^\prime \ = \ \frac{Q}{\sqrt{2}}\left[r^{-4}dx^0\wedge \ldots \wedge dx^3 \wedge dr + d z\wedge {\text{vol}}_{S^4}\right] \ .
\eeq
While the naive kinetic term of the resulting self--dual form vanishes identically, one can argue the normalizability of this perturbation focusing on its electric component.

The superposition of products ${\cal B}_{1,\ell} {\cal B}_{2,m_\ell}$, as in eq.~\eqref{eq:tensor_B1B2}, with $C_\ell\propto\frac{\pi}{2 z_m}$, is an explicit example of $z$--dependent linearized solution that reduces to eq.~\eqref{eq:form_2_variable_10D} at large $r$-distances. At small distances and not close to the endpoints, $H$ and $\widetilde{H}$ in eqs.~\eqref{eq:H_and_H_tilde} are both non-vanishing and the form field strength in eq.~\eqref{eq:two_variable_ansatz_p_brane} approaches the expected ten--dimensional behavior
\beq
{\cal H}_5 \propto \frac{1}{\left(r^2+z^2\right)^\frac{5}{2}} \, dx^0\wedge\ldots dx^3\wedge d\sqrt{r^2 + z^2} \ .
\eeq

\subsubsection{\sc Heterotic SO(16) $\times$ SO(16) Model}

For the heterotic SO(16) $\times$ SO(16) model,
\beq
\beta_p \ = \ \frac{3-p}{4} \ , \label{betap_het}
\eeq
and the limiting behaviors in eq.~\eqref{omegaphi_heterotic} imply that
\beq
\mu \ = \ \frac{\left|p\,-\,1\right|}{4} \ , \qquad \tilde{\mu} \ = \ \frac{\left|p\,-\,5\right|}{4} \ .
\eeq
Consequently, in this case the ratio $\frac{C_1}{C_2}$ parametrizes the allowed boundary conditions at the origin for $p=1$, while the ratio $\frac{C_3}{C_4}$ parametrizes the allowed boundary conditions at the other end for $p=5$. The solution of ${\cal A}\psi=0$ is now, as in eq.~\eqref{eq:form_psi_zeromode},
\bea
p=1 \qquad \psi_1 &=& e^{-\,\frac{1}{2}\, \phi \ +\ 2\,\Omega} \ , \nonumber \\
p=5 \qquad \psi_5 &=& e^{ \frac{1}{2}\, \phi \  - \ 2\,\Omega}  \ .
\eea
Making use of eqs.~\eqref{omegaphi_heterotic}, one can conclude that both $\psi_1$ and $\psi_5$ are not normalizable, since $\psi_1 \sim z^{\frac{1}{2}}$ and $\psi_1 \sim \left(z_m-z\right)^{-\,\frac{1}{2}}$, while $\psi_5 \sim z^{-\, \frac{1}{2}}$ and $\psi_5 \sim \left(z_m-z\right)^{\frac{1}{2}}$ near the two ends of the interval. The Wronskian method yields normalizable solutions along the lines of eq.~\eqref{wrosnk_zm}. In conclusion, the zero modes for the heterotic case are as follows:
\bea
p=1 : \qquad \quad {\cal B} &=&  \frac{Q_1}{r^{5}} \ \int_z^{z_m}  {dz'}\ e^{-\,4\,\Omega(z')\,+\,\phi(z')} \ , \nonumber \\
p=5 : \qquad \quad  {\cal B} &=&  \frac{Q_5}{r} \  \int_0^{z}  {dz'}\ e^{4\,\Omega(z')\,-\,\phi(z')}  \ ,
\eea
which determine the corresponding field strength profiles according to eqs.~\eqref{eq:H_and_H_tilde} and~\eqref{eq:two_variable_ansatz_p_brane},
\bea
{\cal H}_3 &=& -\ \frac{Q_1}{r^5}\left[\frac{5\, dr}{r}\int_z^{z_m}  {dz'}\ e^{-\,4\,\Omega(z')\,+\,\phi(z')} \ + \ dz  \ e^{-\,4\,\Omega(z)\,+\,\phi(z)}\right] \wedge dx^0\wedge dx^1 \ ,  \nonumber\\
{\cal H}_7 &=&  \frac{Q_5}{r} \  \left[- \,\frac{dr}{r}\int_0^{z}  {dz'}\ e^{4\,\Omega(z')\,-\,\phi(z')} \ + \ dz  \ e^{4\,\Omega(z)\,-\,\phi(z)}\right] \wedge dx^0\wedge\ldots \wedge dx^5 \ .
\eea

\subsubsection{\sc On the Back-reaction of Tensor Perturbations}

The next issue that we would like to address is the extent of the back-reaction of these tensor perturbations, at least for sufficiently large values of $r$, where the linearized approximation is supposed to hold. To this end, one can compare the background dilaton contribution to the energy--momentum tensor, which results from $e^{-2
\Omega}\,\phi_{z}^2$ and from the tension term, proportional to $e^{\gamma\phi}$, to the corresponding zero--mode tensor contribution, which is proportional to $e^{-2(p+2)\Omega - 2\beta_p \,\phi}\, \left({\cal B}_r^2+{\cal B}_z^2 \right)$. Making use of the preceding results, one can conclude that, in all cases of interest aside from orientifold D1 brane, the tensor contribution is always subdominant, even at the singular ends. For the orientifold D1 brane, however, the most singular dilaton contribution, which is independent of $r$, can only dominate near the left end of the interval for sufficiently large values of $r$, since
\bea
e^{-2 \Omega}\, \left(\phi_z\right)^2 &\sim&  \left(\frac{z}{z_m}\right)^{-\frac{9}{4}} \ ,  \nonumber \\
e^{-6\,\Omega + \phi} \left({\cal B}_z\right)^2 &\sim& \frac{1}{r^{10}}\ \left(\frac{z}{z_m}\right)^{-\frac{13}{4}}\ ,
\eea
so that the condition translates into
\beq \label{eq:orientifold_p=5_r_ineq}
 r^{10} \ > \ \frac{z_m}{z} \ .
\eeq
In all other cases, it suffices that $r$ be larger than the scale determined by tension for the condition to hold, independently of $z$.

\subsection{\sc Wrapped Charged Branes}

We can now address the asymptotic behavior of branes of the ${\cal H}_{p+3}$ type, which are wrapped in the internal $z$-space.
The corresponding $p+3$ form also decouples from the other perturbations at the linear level, and therefore its field strength is simply given by eq.~\eqref{eq:p+3_form},
\beq
{\cal H}_{p+3} \ = \ \frac{h_0}{r^{7-p}} \ e^{2\beta_{p+1}\phi - (7-p)\Omega} \, \theta^0 \dots \theta^p  \,\theta^r \theta^z \ , \label{eq:hp3}
\eeq
where $\Omega$ and $\phi$ are the background fields. Comparing again with the most singular dilaton contribution to the energy--momentum tensor due to the background reveals that in this case all tensor perturbations are sub-dominant for large values of $r$, with only one exception. This presents itself if $p=0$ for the orientifolds, close to the left end of the interval, where the additional condition
\beq \label{eq:orientifold_p=0_r_ineq}
r^{14} \ > \ \frac{z_m}{z}
\eeq
must hold.
As usual, the orientifold field strength for $p=2$ should be completed into a self-dual expression, which reads
\beq
{\cal H}_5  \ = \ \frac{h_0}{\sqrt{2}} \left[\frac{1}{r^5}\,dx^0 \wedge dx^1 \wedge  dx^2 \wedge dr \wedge dz \ + \ {\text{vol}}_{S^5}\right] \ .
\eeq

\section[Conclusions]{\sc Conclusions}\label{sec:Conclusions}

The core of the present paper was a detailed analysis of the asymptotics of charged and uncharged brane solutions, relying on the linearized approximation for the non--linear system of eqs.~\eqref{eq:2_var_form},~\eqref{eq:p+3_form},~\eqref{eq:2_var_dilaton} and~\eqref{eq:2_var_metric1}--\eqref{eq:2_var_metric5}. The linearized equations mix in a non--trivial fashion contributions from different metric components and the dilaton. They include a normalizable zero mode, a metric perturbation that is independent of the internal coordinate $z$ and has a power--like behavior in $r$, together with an infinite set of modes that decay exponentially for large values of $r$. The zero mode of the metric perturbations captures the dominant large-$r$ behavior of both charged and uncharged branes, while
for the dilaton there are only contributions that decay exponentially for large values of $r$.
We have also seen that, for uncharged branes extended along the internal interval, the perturbations can be completed into exact solutions of the full non-linear system. Perturbations of the form fields are decoupled from the others at the linear level, and allow zero modes with power--like behavior in $r$ (which typically also depend on $z$).

The energy--momentum tensor of form perturbations is generally subdominant with respect to the background contributions, aside from two special cases. These are the $p=1$ orientifold perturbations of the ${
\cal H}_{p+2}$ type, which are only sub-dominant near the left end of the $z$ interval if $r$ satisfies the inequality of eq.~\eqref{eq:orientifold_p=5_r_ineq}, and the $p=0$ orientifold perturbations of the ${
\cal H}_{p+3}$ type, which are only sub--dominant near the left end of the interval if the inequality of eq.~\eqref{eq:orientifold_p=0_r_ineq} holds.

We have seen that the different types of perturbations are in one--to--one correspondence with the cases that emerged long ago from the CFT analysis of~\cite{dms_cft}, where the tadpole potentials were not taken into account. The preceding results are also compliant with the analysis in~\cite{Basile:2022ypo,ms23_1}, which revealed the presence of a mass gap in the dilaton spectrum.

In the only case where the leading contribution is simpler and independent of $z$, which concerns the D3 brane, we also performed some steps aimed at identifying the first corrections to the large--$r$ behavior induced by the tensorial charges. This can be identified expanding the non--linear system of eqs.~\eqref{eq:2_var_metric1}--\eqref{eq:2_var_metric5} in powers of $Q$, and the leading contribution is then of order $Q^2$, as in the familiar case of Reissner--Nordstrom black holes. We leave a precise determination of these terms for future work, but a preliminary analysis indicates that, close to the endpoints of the internal interval and for large distances, the leading corrections introduce, in the background metric, $r$-dependent shifts of the $z$ variable that are proportional to $Q^2$. For instance, close to $z=0$ these corrections can be captured by deforming the background conformal factor $\Omega(z)$ according to
\beq
\Omega\left(z\right)\ \to \  \Omega\left(z \ + \ \frac{a \,z_m \, Q^2}{r^{6}}\right) \ ,
\eeq
with $a$ a constant, and a similar shift emerges near the other endpoint. While these results are suggestive and conform to known black--hole examples, the solution of the linearized system away from the endpoints presents some intricacies, to which we hope to return in a future publication.

\section*{\sc Acknowledgments}
\vskip 12pt
We are grateful to Emilian Dudas for stimulating discussions. AS was supported in part by Scuola Normale and by INFN (IS GSS-Pi). JM is grateful to Scuola Normale Superiore for the kind hospitality while this work was in progress. AS is grateful to Universit\'e  Paris Cit\'e for the kind hospitality, when works along these lines started.

\newpage
\begin{appendices}

\section[Details on the Brane Equations of Section~\ref{sec:branes_with_bulk_tadpoles}]{\sc Details on the Brane Equations of Section~\ref{sec:branes_with_bulk_tadpoles}} \label{app:brane_details}

In this Appendix we collect some details on the derivation of eqs.~\eqref{eq:2_var_form},~\eqref{eq:p+3_form},~\eqref{eq:2_var_dilaton} and~\eqref{eq:2_var_metric1}--\eqref{eq:2_var_metric5}. Here we allow for a more general ansatz,
\bea\label{eq:metric_D_dim}
ds^2 &=& e^{2A(\zeta,u)}dx_{p+1}^2 \, + \, e^{2B(\zeta,u)} \Big(d u \,+\, E(\zeta,u) \,d\zeta \Big)^2 \, + \, e^{2D(\zeta,u)} d\zeta^2 \,+\, e^{2C(\zeta,u)} d\,\Omega^2_{{\cal D}-p-3}  \ , \nonumber \\
\phi &=& \phi(\zeta,u) \ ,
\eea
while also working in a general dimension ${\cal D}$,
so that $(z,r)$ are arbitrary functions of $(\zeta,u)$, and a two--parameter gauge symmetry is still present at this stage. As in Section~\ref{sec:branes_with_bulk_tadpoles}, either of the two field strengths
\beq
{\cal H}_{p+2} \ = \  {\cal H}_{p+2}(\zeta,u) \quad \text{or} \quad {\cal H}_{p+3} \ = \  {\cal H}_{p+3}(\zeta,u)
\eeq
is also present,
depending on whether one is considering a $p$-brane or a $(p+1)$-brane whose world-volume is extended along the $\zeta$ direction. In special cases, when one is a NS brane and the other is a type-IIA brane, one could also consider the two contributions simultaneously, but this option does not concern the ten--dimensional non--supersymmetric non--tachyonic strings, which are the main target of our analysis.
These backgrounds rest on a number of functions of two variables that can capture the behavior of brane solutions in the presence of the tadpole potential.

\subsection{\sc Derivation of the Non-Linear Equations}

The moving basis corresponding to the metric of eq.~\eqref{eq:metric_D_dim} is
\bea
\theta^\mu\ = \ e^A\, dx^\mu\,,\ \ \ \theta^u\ = \ e^B\left(du\ + \ E \,d\zeta\right)\,, \ \ \ \theta^\zeta\ = \ e^D d\zeta,\ \ \ \theta^a\ = \ e^C\tilde{\theta}^a\, ,
\eea
where the one-forms $\tilde{\theta}^a$ refer to the unit sphere and with $\mu=0,\dots p$, $a=1,\dots {\cal D}-p-3$. This setup leads to the curvature two-forms
\bea
\Omega_{\mu}{}^{\nu}&=&-\ \theta_\mu\ \theta^\nu\ [e^{-2B}\ A_u^2 \ + \ e^{-2D}( A_\zeta-EA_u)^2]\, , \nonumber \\
\Omega_u{}^\mu&=&\theta^u\ \theta^\mu\left[e^{-2D}(A_\zeta-EA_u)(EB_u+E_u-B_\zeta) \ - \ e^{-A-B}(e^{A-B}A_u)_u\right]\nonumber\\
&+&\theta^\zeta\ \theta^\mu\left[
e^{-A-D} \left(
E\left(e^{A-B}A_u\right)_u-\left(e^{A-B}A_u\right)_\zeta
\right)
\ + \ e^{-D-B}\ D_u\ (A_\zeta-EA_u)
\right]\,, \nonumber\\
\Omega _\zeta{}^\mu &=& -\ \theta^u\ \theta^\mu
\left[
e^{-A-B}
\left(e^{A-D}(A_\zeta-E A_u)\right)_u
\ + \ e^{-B-D}\ A_u\ (E B_u + E_u - B_\zeta)
\right]\nonumber\\
&+& \theta^\zeta\ \theta^\mu \left[
e^{-A-D}E \left(e^{A-D}(A_\zeta-E A_u)\right)_u \nonumber \right. \\
&-& \left. e^{-A-D}\left(e^{A-D}(A_\zeta-E A_u)\right)_\zeta\ - \ e^{-2B}A_u\ D_u
\right] \,, \nonumber\\
\Omega_a{}^{\mu}&=& \theta^\mu\ \theta_a\left[
e^{-2B}A_uC_u\ + \ e^{-2D}(A_\zeta-EA_u)(C_\zeta-EC_u)
\right]\,, \nonumber\\
\Omega_u{}^\zeta &=&\theta^u \theta^\zeta\ \left[
e^{-B-D}\left(e^{B-D}\left( E B_u+E_u-B_\zeta\right)
\right)_\zeta \right. \nonumber \\
&-& \left. e^{-B-D}\left(e^{B-D}E \left( E B_u+E_u-B_\zeta\right)
\ +\ e^{D-B}D_u\right)_u
\right]\,,\nonumber\\
\Omega_u{}^a&=&\theta^u \theta^a \ \left[
e^{-2D}(EB_u+E_u-B_\zeta)(C_\zeta - E C_u)\ - \ e^{-B-C}\left( e^{C-B}C_u\right)_u
\right]\nonumber\\
&+&\theta^\zeta \theta^a \left[ E e^{-C-D}\left( e^{C-B}C_u\right)_u-e^{-C-D}\left( e^{C-B}C_u\right)_\zeta
\ + \ e^{-B-D}D_u(C_\zeta-EC_u)
\right]\,, \nonumber\\
\Omega_\zeta{}^a&=& - \, \theta^u \theta^a \left[
e^{-B-D}C_u(EB_u+E_u-B_\zeta)\ + \ e^{-B-C}\left(e^{C-D}(C_\zeta-EC_u)
\right)_u\right]\nonumber\\
&+&\theta^\zeta \theta^a \left[
E e^{-D-C}\left(e^{C-D}\left(C_\zeta-E C_u\right)\right)_u \ - \ e^{-D-C}\left(e^{C-D}\left(C_\zeta-E C_u\right)\right)_\zeta \nonumber \right. \\ &-& \left.
e^{-2B} D_u C_u \right]\,, \nonumber  \\
\Omega_a{}^b&=&\tilde\Omega_a{}^b\ - \ \theta_a \theta^b\left[e^{-2B}(C_u)^2\ + \ e^{-2D}(C_\zeta-EC_u)^2\right]\,,
\eea
where in the last expression $\tilde\Omega_a{}^b$ denotes the curvature of the unit sphere.

From the general decomposition
\bea
\Omega_A{}^B\ = \ \frac{1}{2}\ \Omega_{CDA}{}^B\,\theta^C \theta^D \ ,
\eea
one can deduce the flat components of the Riemann tensor,
\bea
\Omega_{\alpha\beta\mu}{}^{\nu}&=&\left(\eta_{\beta\mu}\ \delta_\alpha^\nu -\eta_{\alpha\mu}\ \delta_\beta^\nu\ \right)\left[e^{-2B}\ A_u^2+ e^{-2D}( A_\zeta-EA_u)^2\right] ,\nonumber \\
\Omega_{u\alpha u}{}^\mu&=&\ \delta_\alpha^\mu\left[e^{-2D}(A_\zeta-EA_u)(EB_u+E_u-B_\zeta)-e^{-A-B}(e^{A-B}A_u)_u\right] , \nonumber\\
\Omega_{\zeta\alpha u}{}^\mu&=& \delta_\alpha^\mu\left[
e^{-A-D} \left(
E\left(e^{A-B}A_u\right)_u-\left(e^{A-B}A_u\right)_\zeta
\right)
+e^{-D-B}\ D_u\ (A_\zeta-EA_u)
\right],  \nonumber \\
\Omega_{u\alpha \zeta}{}^\mu&=& -\ \delta_\alpha^\mu
\left[e^{-A-B}\left(e^{A-D}(A_\zeta-E A_u)\right)_u + e^{-B-D}\ A_u\ (E B_u + E_u - B_\zeta)\right], \nonumber\\
\Omega_{\zeta \alpha \zeta}{}^\mu &=&  \delta_\alpha^\mu \,e^{-A-D}\left[
E \left(e^{A-D}(A_\zeta-E A_u)\right)_u
-\left(e^{A-D}(A_\zeta-E A_u)\right)_\zeta\ - \ e^{A+D-2B}A_u\ D_u
\right] , \nonumber\\
\Omega_{\alpha b a}{}^{\mu}&=& \delta_\alpha^\mu\ \delta_{ba}\left[ e^{-2B}A_uC_u+e^{-2D}(A_\zeta-EA_u)(C_\zeta-EC_u)
\right],\nonumber \\
\Omega_{u\zeta u}{}^\zeta &=& e^{-B-D} \left[
\left(e^{B-D}\left( E B_u+E_u-B_\zeta\right)
\right)_\zeta
-\left(e^{B-D}E \left( E B_u+E_u-B_\zeta\right)
\ +\ e^{D-B}D_u\right)_u
\right],\nonumber\\
\Omega_{u b u}{}^a&=& \delta_b^a \ \left[
e^{-2D}(EB_u+E_u-B_\zeta)(C_\zeta - E C_u)-e^{-B-C}\left( e^{C-B}C_u\right)_u
\right], \nonumber\\
\Omega_{\zeta b u}{}^a&=&\delta_b^a \,e^{-C-D}\left[ E \left( e^{C-B}C_u\right)_u-\left( e^{C-B}C_u\right)_\zeta
+e^{C-B}D_u(C_\zeta-EC_u)
\right], \nonumber\\
\Omega_{u b \zeta}{}^a&=&- \ \delta_b^a \left[
e^{-B-D}C_u(EB_u+E_u-B_\zeta)+e^{-B-C}\left(e^{C-D}(C_\zeta-EC_u)
\right)_u\right],\nonumber\\
\Omega_{\zeta b \zeta}{}^{a}&=&\delta_b^a \,e^{-C-D}\left[
E \left(e^{C-D}(C_\zeta-E C_u)\right)_u-\left(e^{C-D}\left(C_\zeta-E C_u\right)\right)_\zeta \, - \,
e^{C+D-2B} D_u C_u \right],\nonumber\\
\Omega_{cda}{}^b&=&e^{-2C} \, \tilde\Omega_{cda}{}^b \ - \ \left(\delta_{ca} \delta_d^b\,-\,\delta_{da}\delta_c^b\right)\left[e^{-2B}(C_u)^2 \ + \ e^{-2D}(C_\zeta-EC_u)^2\right]\,.
\eea
The other components vanish, or can be deduced from these by antisymmetry.
One can then deduce the Ricci tensor in the moving basis,
\bea
R_{AB}\ = \ \Omega_{ADB}{}^D\ ,
\eea
and its  non--vanishing components are
\begin{eqnarray}
R_{\alpha\mu}&=&\Omega_{\alpha\nu\mu}{}^\nu\ + \ \Omega_{\alpha u\mu}{}^u \ + \ \Omega_{\alpha\zeta\mu}{}^\zeta\ + \ \Omega_{\alpha a\mu}{}^a\ , \nonumber \\
R_{uu}&=&\Omega_{u \nu u }{}^\nu\  + \ \Omega_{u\zeta u}{}^\zeta\ + \ \Omega_{u a u}{}^a \ , \nonumber \\
R_{u\zeta}&=&\Omega_{u \nu \zeta }{}^\nu \   + \ \Omega_{u a \zeta}{}^a\ , \nonumber \\
R_{\zeta\zeta}&=&\Omega_{\zeta \nu \zeta }{}^\nu \ + \ \Omega_{\zeta u \zeta}{}^u \  + \ \Omega_{\zeta a \zeta}{}^a \ , \nonumber \\
R_{c a }&=&\Omega_{c\nu a}{}^\nu \ + \ \Omega_{c u a}{}^u \ + \ \Omega_{c\zeta a}{}^\zeta \ + \ \Omega_{c b a}{}^b \ .
\end{eqnarray}
In detail, one finds
\bea
R_{\alpha\mu} &=& - \ \eta_{\alpha\mu}\Big\{ e^{-2B} \left(A_{uu}+A_u F^+_u\right) + e^{-2D}\left[\nabla_\zeta\nabla_\zeta A +\nabla_\zeta A \left(\nabla_\zeta F^- - E_u\right)\right]\!\!\Big\} \, ,  \nonumber  \\
R_{uu} &=& - \ e^{-2B} \Big[(F^+ +B)_{uu} +(p+1)A_u^2- B_u^2 +({\cal D}-p-3)C_u^2+D_u^2-B_u F^+_u\Big] \nonumber \\
    &-& e^{-2D}\Big[ \nabla_\zeta \left(\nabla_\zeta B-E_u\right)+(\nabla_\zeta B-E_u)\nabla_\zeta F^- \Big] \, ,  \nonumber \\
R_{u\zeta} &=&\frac{1}{2} \ e^{-B-D}\Big\{-\left[ \nabla_\zeta \left(F^+ + F^-\right)\right]_u + D_u \nabla_\zeta \left(F^+ + F^-\right)   \\
    &+& \left(\nabla_\zeta B - E_u\right)\left(F^+ + F^-\right)_u - 2(p+1) A_u\nabla_\zeta A-2({\cal D}-p-3) C_u\nabla_\zeta C \Big\} \, ,  \nonumber  \\
R_{\zeta\zeta} &=& - \ e^{-2B}\Big( D_{uu}+D_u F^+_u \Big) - e^{-2D}\Big[\nabla_\zeta\nabla_\zeta (F^- + D) + (p+1)(\nabla_\zeta A)^2 + (\nabla_\zeta B)^2  \nonumber \\
    &+&({\cal D}-p-3)(\nabla_\zeta C)^2- (\nabla_\zeta D)^2- \nabla_\zeta D \nabla_\zeta F^- - \nabla_\zeta E_u - E_u \nabla_\zeta (B-D)\Big]\, ,  \nonumber  \\
R_{ac} &=& e^{-2C} \, \tilde{R}_{ac}  -  \delta_{ac}\Big\{e^{-2B}\left(C_{uu}+C_u F^+_u\right)+e^{-2D}\left[\nabla_\zeta\nabla_\zeta C+\nabla_\zeta C \left(\nabla_\zeta F^- -E_u\right)\right]\Big\} \, , \nonumber
\eea
where we used
\beq
\nabla_\zeta\ =\ \partial_\zeta \ - \ E \, \partial_u \ ,
\eeq
and
\beq
F^{\pm} \ = \ (p+1)A \ \mp\  B\ +\ ({\cal D}-p-3)C \ \pm\  D \ .
\eeq
Consequently, the Ricci scalar is
\bea
{\cal R}&=& - \ e^{-2B}\Big[ 2\left(F^+ + B\right)_{uu} +(p+1)A_u^2- B_u^2 +({\cal D}-p-3)C_u^2+D_u^2+ (F^+_u)^2\Big] \nonumber \\
&+& e^{-2C} \, \tilde{R} \ -\ e^{-2D}\Big[ 2\nabla_\zeta\nabla_\zeta (F^- + D) + (p+1)(\nabla_\zeta A)^2 + (\nabla_\zeta B)^2 \nonumber \\
        &+& ({\cal D}-p-3)(\nabla_\zeta C)^2 -(\nabla_\zeta D)^2+(\nabla_\zeta F^-)^2 - 2 E_u \nabla_\zeta F^- -2 \nabla_\zeta E_u\Big]\, .
\eea

The Ricci tensor in the coordinate basis is then obtained using
\bea
{\cal R}_{MN}\ = \ R_{AB}\ \theta_M{}^A\ \theta_N{}^B \ ,
\eea
where the $\theta_M{}^A$ can be extracted from the moving basis according to
\bea
\theta^A\ = \ \theta_M{}^A\ dx^M \ .
\eea
One thus obtains
\bea{\cal R}_{\mu\nu} &=& e^{2A} R_{\mu\nu} \ , \qquad {\cal R}_{uu} \ = \ e^{2B} R_{uu} \ , \qquad {\cal R}_{u\zeta} \ = \ e^{B+D}R_{u\zeta}\ + \ E e^{2B}R_{uu} \ , \nonumber \\
{\cal R}_{\zeta\zeta} &=& e^{2D}R_{\zeta\zeta}\ + \ 2\, E\, e^{B+D}R_{u\zeta} \ + \ E^2\, e^{2B} \,R_{uu} \ , \qquad {\cal R}_{ab} \ = \ e^{2C} R_{ab} \ .
\eea
Finally, the spacetime components of the Einstein tensor are
\bea
{\cal G}_{\mu\nu} &=& - \ \frac{1}{2} \ \eta_{\mu\nu}\Big\{ e^{2(A-B)}\Big[ 2 \left(A-B- F^+\right)_{uu} +2 A_u F^+_u -  (p+1)A_u^2+ B_u^2 \nonumber  \\
    &-&({\cal D}-p-3)C_u^2-D_u^2 - (F^+_u)^2 \Big]  \ + \ e^{2(A-C)} \, \tilde{R} \nonumber \\
        &+& e^{2(A-D)}\Big[ 2\nabla_\zeta\nabla_\zeta (A-D-F^- ) +2\nabla_\zeta A \nabla_\zeta F^- - (p+1)(\nabla_\zeta A)^2 - (\nabla_\zeta B)^2 \nonumber \\
            &-&({\cal D}-p-3)(\nabla_\zeta C)^2 +(\nabla_\zeta D)^2-(\nabla_\zeta F^-)^2 + 2 E_u \nabla_\zeta (F^- -A)+2 \nabla_\zeta E_u \Big]\Big\} \, , \nonumber  \\
{\cal G}_{uu} &=& - \ \frac{1}{2} \ \Big\{ \Big[(p+1)A_u^2- (B_u + F^+_u)^2 +({\cal D}-p-3)C_u^2+D_u^2\Big] +e^{2(B-C)} \, \tilde{R}\nonumber \\
    &+&  e^{2(B-D)}\Big[2 \nabla_\zeta \nabla_\zeta (B-D-F^-) -(p+1)(\nabla_\zeta A)^2 -(\nabla_\zeta B-\nabla_\zeta F^-)^2 \nonumber \\
        &-&  ({\cal D}-p-3)(\nabla_\zeta C)^2 +(\nabla_\zeta D)^2\Big]\Big\} \, ,\nonumber \\
{\cal G}_{u\zeta} &=&\frac{1}{2} \ \Big\{-\left[ \nabla_\zeta \left(F^+ + F^-\right)\right]_u + D_u \nabla_\zeta \left(F^+ + F^-\right)\nonumber   \\
    &+& \left(\nabla_\zeta B - E_u\right)\left(F^+ + F^-\right)_u - 2(p+1) A_u\nabla_\zeta A-2({\cal D}-p-3) C_u\nabla_\zeta C \Big\} \nonumber \\
        &-&  \frac{1}{2} \ E \ \Big\{(p+1)A_u^2- (B_u+F^+_u)^2 +({\cal D}-p-3)C_u^2+D_u^2 + e^{2(B-C)} \, \tilde{R} \nonumber \\
            &+&  e^{2(B-D)}\Big[2 \nabla_\zeta\nabla_\zeta (B-D-F^-)-(p+1)(\nabla_\zeta A)^2- (\nabla_\zeta B - \nabla_\zeta F^-)^2 \nonumber \\
                &-& ({\cal D}-p-3)(\nabla_\zeta C)^2 +(\nabla_\zeta D)^2 \Big]\Big\} \, , \\
{\cal G}_{\zeta\zeta} &=& - \ \frac{1}{2} \ \Big\{ \Big[ (p+1)(\nabla_\zeta A)^2 + (\nabla_\zeta B)^2 +({\cal D}-p-3)(\nabla_\zeta C)^2 -(\nabla_\zeta D + \nabla_\zeta F^-)^2\nonumber \\
        &-&  2 E_u \nabla_\zeta (B-D-F^-)\Big] + e^{2(D-C)} \, \tilde{R} \ + \ e^{2(D-B))}\Big[ 2\left(D-B-F^+\right)_{uu}  \nonumber \\
            &-& (p+1)A_u^2+ B_u^2 -({\cal D}-p-3)C_u^2-(D_u-F^+_u)^2\Big]\Big\} \nonumber \\
                &+& E \ \Big\{-\left[ \nabla_\zeta \left(F^+ + F^-\right)\right]_u + D_u \nabla_\zeta \left(F^+ + F^-\right)\nonumber   \\
                &+& \left(\nabla_\zeta B - E_u\right)\left(F^+ + F^-\right)_u - 2(p+1) A_u\nabla_\zeta A-2({\cal D}-p-3) C_u\nabla_\zeta C \Big\} \nonumber \\
                    &-& E^2 \ \Big[(F^+ +B)_{uu} +(p+1)A_u^2- B_u^2 +({\cal D}-p-3)C_u^2+D_u^2-B_u F^+_u\Big] \nonumber \\
                        &-& E^2 \ e^{2(B-D)}\Big[ \nabla_\zeta \left(\nabla_\zeta B-E_u\right)+(\nabla_\zeta B-E_u)\nabla_\zeta F^- \Big] \, ,\nonumber  \\
{\cal G}_{mn} &=& - \ \frac{1}{2} \tilde{\gamma}_{mn} \Big\{ \frac{{\cal D}-p-5 }{{\cal D}-p-3} \ \tilde{R} \ + e^{2(C-B)}\Big[2(C-B-F^+)_{uu}+2C_u F^+_u-(p+1)A_u^2 \nonumber \\
    &+& B_u^2 - ({\cal D}-p-3)C_u^2 - D_u^2 - (F^+_u)^2 \Big]+ e^{2(C-D)} \Big[2\nabla_\zeta\nabla_\zeta (C-D-F^-) \nonumber  \\
        &+& 2\nabla_\zeta C\nabla_\zeta F^- -(p+1)(\nabla_\zeta A)^2 - (\nabla_\zeta B)^2 - ({\cal D}-p-3)(\nabla_\zeta C)^2+(\nabla_\zeta D)^2\nonumber \\
                &-& (\nabla_\zeta F^-)^2 + 2 E_u \nabla_\zeta (F^- - C) +2 \nabla_\zeta E_u \Big] \Big\}\nonumber  \, ,
\eea
where $\tilde{\gamma}_{mn}$ is the metric on the round unit sphere.

A general expression for the $(p+2)$-form field strength that is compatible with the symmetries of the background rests on a pair of functions, $H(\zeta,u)$ and $\widetilde{H}(\zeta,u)$, and reads
\beq
{\cal H}_{p+2} \ = \ H(\zeta,u) \, \theta^u \, \theta^0 \ldots \theta^p  \ + \ \widetilde{H}(\zeta,u) \,\theta^\zeta \,\theta^0 \ldots \theta^p  \ .
\eeq
The Bianchi identity and the equations of motion translate into a pair of first--order equations linking $H$ and $\widetilde{H}$:
\bea
\left[H \ e^{(p+1)A+B}\right]_\zeta \!\!&=&\!\! \left[H\ E \  e^{(p+1)A+B} \ + \ \widetilde{H} \ e^{(p+1)A+D}\right]_u \ ,  \\
\left[\widetilde{H} \ e^{-2\beta_p\phi + B + ({\cal D}-p-3)C}\right]_\zeta \!\!&=&\!\! \left[\widetilde{H} \ E \  e^{-2\beta_p\phi + B + ({\cal D}-p-3)C} \ - \  H \  e^{-2\beta_p\phi + ({\cal D}-p-3)C + D}\right]_u  \nonumber \ .
\eea
One can solve the Bianchi identity by introducing the potential
\beq
{\cal B}_{p+1} \ = \ e^{-(p+1)A}  \ {\cal B}(\zeta,u) \, \theta^0 \ldots \theta^p \ ,
\eeq
so that $H(\zeta,u)$ and $\widetilde{H}(\zeta,u)$ can be expressed in terms of ${\cal B}$ as
\beq
H \ = \  e^{-(p+1)A-B} \ {\cal B}_u \ , \qquad
\widetilde{H} \ = \ e^{-(p+1)A-D} \ \nabla_\zeta {\cal B} \ .
\eeq
In terms of ${\cal B}$, the form equation reads
\beq \label{eq:Beq}
\nabla_\zeta\left( e^{2G} \ \nabla_\zeta {\cal B} \right)\ = \ E_u \ e^{2G} \ \nabla_\zeta {\cal B}  \ - \ \left[e^{2(G -B  + D)} \ {\cal B}_u  \right]_u \ ,
\eeq
where we defined
\beq
2G \ = \ - \ 2\beta_p\phi \ - \ (p+1)A \ +\  B \ +\  ({\cal D}-p-3)C\  - \ D \ .
\eeq
The contributions of this $(p+2)$-form to the equations of motion involve $\left({\cal H}_{p+2}^2\right)_{MN}$, defined as
\beq
\left({\cal H}_{p+2}^2\right)_{MN} \ = \ {\left({\cal H}_{p+2}\right)}_{M M_2 \ldots M_{p+2}}\,{\left({\cal H}_{p+2}\right)}_{N N_2 \ldots N_{p+2}} \  g^{M_2 N_2}\ldots g^{M_{p+2} N_{p+2}}\ .
\eeq
Its non--vanishing components read
\bea
\frac{\left({{\cal H}_{p+2}^2}\right)_{\mu\nu}}{(p+1)!} &=& - \  \eta_{\mu\nu} \  e^{2A} \left(H^2 \ + \ \widetilde{H}^2\right) \ ,  \ \
\frac{\left({{\cal H}_{p+2}^2}\right)_{uu}}{(p+1)!} \ = \ - \ e^{2B}\  H^2 \ ,  \\
\frac{\left({{\cal H}_{p+2}^2}\right)_{u\zeta}}{(p+1)!} &=& - \  e^{B+D} \  H \ \widetilde{H}  \ - \  E  \ e^{2B} \ H^2  \ , \ \
\frac{\left({{\cal H}_{p+2}^2}\right)_{\zeta\zeta}}{(p+1)!} \ = \  - \left(e^D \ \widetilde{H} \ + \ E \ e^B \ H \right)^2 \ , \nonumber
\eea
and the resulting trace is
\beq
\frac{{\cal H}_{p+2}^2}{(p+2)!} \ = \  -\  \left(H^2 \ + \ \widetilde{H}^2\right) \ .
\eeq

The other case of interest is the $(p+3)$-form field strength
\beq
{\cal H}_{p+3} \ = \ h(\zeta,u) \, \theta^0 \dots \theta^p  \,\theta^u \, \theta^\zeta \ ,
\eeq
whose Bianchi identity is automatically satisfied, and the corresponding equations are solved by
\beq
h(\zeta,u) \ = \ h_0 \ e^{2\beta_{p+1}\phi - ({\cal D}-p-3)C} \ ,
\eeq
where $h_0$ is a constant. This leads to
\bea
\frac{\left({{\cal H}_{p+3}^2}\right)_{\mu\nu}}{(p+2)!} &=& - \ \eta_{\mu\nu}\ e^{2A} \ h^2(\zeta,u) \ , \qquad \frac{\left({{\cal H}_{p+3}^2}\right)_{uu}}{(p+2)!} \ = \ - \ e^{2B}\  h^2(\zeta,u) \ , \nonumber \\
\frac{\left({{\cal H}_{p+3}^2}\right)_{u\zeta}}{(p+2)!} &=&  - \  E  \ e^{2B} \ h^2(\zeta,u)  \ , \qquad \frac{\left({{\cal H}_{p+3}^2}\right)_{\zeta\zeta}}{(p+2)!}  \ = \ - \left( e^{2D}  \ + \ E^2 \ e^{2B}\right) h^2(\zeta,u) \ , \nonumber \\
\frac{{\cal H}_{p+3}^2}{(p+3)!} &=& - \ h^2(\zeta,u) \ .
\eea

The metric equations are now
\bea
\mathcal{G}_{MN} &=& \frac{4}{{\cal D}-2}\  \pr_M\phi\, \pr_N\phi\, + \, \frac{e^{\,-\,2\,\beta_p\,\phi} }{2(p+1)!}\,  \left({\cal H}_{p+2}^2\right)_{M N} \, + \, \frac{e^{\,-\,2\,\beta_{p+1}\,\phi} }{2(p+2)!}\,  \left({\cal H}_{p+3}^2\right)_{M N}  \nonumber\\
&-& \frac{1}{2}\,g_{MN}\left[\frac{4\,(\pr\phi)^2}{{\cal D}-2}+ \frac{e^{\,-\,2\,\beta_p\,\phi}}{2(p+2)!}\,{\cal H}_{p+2}^2\,+ \frac{e^{\,-\,2\,\beta_{p+1}\,\phi}}{2(p+3)!}\,{\cal H}_{p+3}^2\,+\, V(\phi)\right] \ .
\eea
In detail, the $\mu\nu$ components lead to
\bea
\mu\nu&:& e^{-2B} \Big[ 2 \left(A-B- F^+\right)_{uu} +2 A_u F^+_u -  (p+1)A_u^2+ B_u^2 \nonumber  \\
    &-&({\cal D}-p-3)C_u^2-D_u^2 - (F^+_u)^2 \Big]  \ + \ e^{-2C} \, \tilde{R} \nonumber \\
        &+& e^{-2D}\Big[ 2\nabla_\zeta\nabla_\zeta (A-D-F^- ) +2\nabla_\zeta A \nabla_\zeta F^- - (p+1)(\nabla_\zeta A)^2 - (\nabla_\zeta B)^2 \nonumber \\
            &-&({\cal D}-p-3)(\nabla_\zeta C)^2 +(\nabla_\zeta D)^2-(\nabla_\zeta F^-)^2 + 2 E_u \nabla_\zeta (F^- -A)+2 \nabla_\zeta E_u \Big] \nonumber \\
                &=& \frac{1}{2}\, \left[e^{-2B} \left({\cal B}_u\right)^2 \, + \,  e^{-2D} \left(\nabla_\zeta {\cal B}\right)^2 \right] e^{\,-\,2\,\beta_p\,\phi\, -\,2\,(p+1)\,A} \, + \, \frac{1}{2} \,  h_0^2 \ e^{2\,\beta_{p+1}\,\phi\,-\,2\,({\cal D}-p-3)\,C} \nonumber\\
        &+& \frac{4}{{\cal D}-2}\left[e^{-2B}\phi_u^2\, + \,  e^{-2D}\left(\nabla_\zeta \phi\right)^2\right] \  + \  T \, e^{\gamma\,\phi} \label{munu_complete}  \ ,
\eea
the $uu$ component is
\bea
uu&:& e^{-2B} \Big[(p+1)A_u^2- (B_u + F^+_u)^2 +({\cal D}-p-3)C_u^2+D_u^2\Big] +e^{-2C} \, \tilde{R}\nonumber \\
    &+&  e^{-2D}\Big[2 \nabla_\zeta \nabla_\zeta (B-D-F^-) -(p+1)(\nabla_\zeta A)^2 -(\nabla_\zeta B-\nabla_\zeta F^-)^2 \nonumber \\
        &-&  ({\cal D}-p-3)(\nabla_\zeta C)^2 +(\nabla_\zeta D)^2\Big]\nonumber \\
        &=&   \frac{4}{{\cal D}-2}\left[- \ e^{-2B}\phi_u^2\ + \  e^{-2D}\left(\nabla_\zeta \phi\right)^2\right] \  +  \ \frac{1}{2} \ h_0^2  \, e^{2\, \beta_{p+1}\,\phi\, -\, 2\,({\cal D}-p-3)\,C}  \nonumber \\
        &+& \frac{1}{2} \left[e^{-2B} \left({\cal B}_u\right)^2 \ - \  e^{-2D} \left(\nabla_\zeta {\cal B}\right)^2 \right] \, e^{-\,2\,\beta_p\,\phi\,-\,2\,(p+1)\,A} \ + \  T \, e^{\gamma\,\phi} \ ,
\eea
the $u\zeta$ component is
\bea
u\zeta &:& -\left[ \nabla_\zeta \left(F^+ + F^-\right)\right]_u + D_u \nabla_\zeta \left(F^+ + F^-\right)\nonumber   \\
    &+& \left(\nabla_\zeta B - E_u\right)\left(F^+ + F^-\right)_u - 2(p+1) A_u\nabla_\zeta A-2({\cal D}-p-3) C_u\nabla_\zeta C  \nonumber \\
        &-&  E \ \Big\{(p+1)A_u^2- (B_u+F^+_u)^2 +({\cal D}-p-3)C_u^2+D_u^2 + e^{2(B-C)} \, \tilde{R} \nonumber \\
            &+&  e^{2(B-D)}\Big[2 \nabla_\zeta\nabla_\zeta (B-D-F^-)-(p+1)(\nabla_\zeta A)^2- (\nabla_\zeta B - \nabla_\zeta F^-)^2 \nonumber \\
                &-& ({\cal D}-p-3)(\nabla_\zeta C)^2 +(\nabla_\zeta D)^2 \Big]\Big\}  \nonumber \\
                    &=& \frac{8}{{\cal D}-2}\ \phi_u\nabla_\zeta\phi\, - \,  e^{-\,2\,\beta_p\,\phi\,-\,2\,(p+1)\,A} \ {\cal B}_u \nabla_\zeta {\cal B}   \nonumber \\
&+& \frac{4E}{{\cal D}-2}\left[\phi_u^2\ - \   e^{2(B-D)}\left(\nabla_\zeta \phi\right)^2\right] \  -  \ \frac{E \ h_0^2}{2}   \, e^{2\,\beta_{p+1}\,\phi\,+\,2\,B\,-\, 2\,({\cal D}-p-3)\,C}  \nonumber \\
        &-& \frac{E}{2} \  \left[\left({\cal B}_u\right)^2 \ - \  e^{2(B-D)} \left(\nabla_\zeta {\cal B}\right)^2 \right] \, e^{-\,2\,\beta_p\,\phi\,-\,2\,(p+1)\,A} \ - \  T \,E \,e^{2B} \, e^{\gamma\,\phi} \ ,
\eea
the $\zeta\zeta$ component is
\bea
\zeta\zeta &:&  \Big[ (p+1)(\nabla_\zeta A)^2 + (\nabla_\zeta B)^2 +({\cal D}-p-3)(\nabla_\zeta C)^2 -(\nabla_\zeta D + \nabla_\zeta F^-)^2\nonumber \\
        &-&  2 E_u \nabla_\zeta (B-D-F^-)\Big] + e^{2(D-C)} \, \tilde{R} \ + \ e^{2(D-B)}\Big[ 2\left(D-B-F^+\right)_{uu}  \nonumber \\
            &-& (p+1)A_u^2+ B_u^2 -({\cal D}-p-3)C_u^2-(D_u-F^+_u)^2\Big] \nonumber \\
                &-& 2\  E \ \Big\{-\left[ \nabla_\zeta \left(F^+ + F^-\right)\right]_u + D_u \nabla_\zeta \left(F^+ + F^-\right)\nonumber   \\
                &+& \left(\nabla_\zeta B - E_u\right)\left(F^+ + F^-\right)_u - 2(p+1) A_u\nabla_\zeta A-2({\cal D}-p-3) C_u\nabla_\zeta C \Big\} \nonumber \\
                    &+& 2 \ E^2 \ \Big[(F^+ +B)_{uu} +(p+1)A_u^2- B_u^2 +({\cal D}-p-3)C_u^2+D_u^2-B_u F^+_u\Big] \nonumber \\
                    &+& 2 \  E^2 \ e^{2(B-D)}\Big[ \nabla_\zeta \left(\nabla_\zeta B-E_u\right)+(\nabla_\zeta B-E_u)\nabla_\zeta F^- \Big] \nonumber \\
                        &=& \frac{4}{{\cal D}-2}\ \left[\left(e^{D-B}\,\phi_u \,-\, E\, e^{B-D} \,\nabla_\zeta\phi\right)^2 \,-\, \left(\nabla_\zeta\phi\, +\, E \,\phi_u\right)^2\right]   \nonumber \\
        &-& \frac{1}{2} \  \left[\left(e^{D-B}\,{\cal B}_u\, -\, E \,e^{B-D}\,\nabla_\zeta{\cal B}\right)^2 \,- \,\left(\nabla_\zeta{\cal B} \,+\, E {\cal B}_u\right)^2
        \right] \, e^{-\,2\,\beta_p\,\phi\,-\,2\,(p+1)\,A} \nonumber \\
        &+& \frac{1}{2} \ (e^{2D}+E^2 e^{2B})\  h_0^2 \, e^{2\, \beta_{p+1}\,\phi \,-\, 2\,({\cal D}-p-3)\,C} \ + \ (e^{2D}+E^2 e^{2B}) \  T \, e^{\gamma\,\phi} \ ,
\eea
and finally the $mn$ components lead to
\bea
mn&:& e^{-2C} \frac{{\cal D}-p-5 }{{\cal D}-p-3} \ \tilde{R} \ + e^{-2B}\Big[2(C-B-F^+)_{uu}+2C_u F^+_u-(p+1)A_u^2 \nonumber \\
    &+& B_u^2 - ({\cal D}-p-3)C_u^2 - D_u^2 - (F^+_u)^2 \Big]+ e^{-2D} \Big[2\nabla_\zeta\nabla_\zeta (C-D-F^-) \nonumber  \\
        &+& 2\nabla_\zeta C\nabla_\zeta F^- -(p+1)(\nabla_\zeta A)^2 - (\nabla_\zeta B)^2 - ({\cal D}-p-3)(\nabla_\zeta C)^2+(\nabla_\zeta D)^2\nonumber \\
                &-& (\nabla_\zeta F^-)^2 + 2 E_u \nabla_\zeta (F^- - C) +2 \nabla_\zeta E_u \Big] \nonumber \\
                    &=& \frac{4}{{\cal D}-2}\left[e^{-2B}\phi_u^2\ + \  e^{-2D}\left(\nabla_\zeta \phi\right)^2\right] \  -  \ \frac{1}{2} \ h_0^2  \, e^{2 \,\beta_{p+1}\,\phi \,-\, 2\,({\cal D}-p-3)\,C}  \nonumber \\
        &-& \frac{1}{2} \ \left[e^{-2B} \left({\cal B}_u\right)^2 \ + \  e^{-2D} \left(\nabla_\zeta {\cal B}\right)^2 \right] \, e^{-\,2\,\beta_p\,\phi\,-\,2\,(p+1)\,A} \ + \  T \, e^{\gamma\,\phi} \ .
\eea
These are the only non--trivial components of the Einstein equations.

Moreover, the dilaton equation is
\bea
 \frac{8}{{\cal D}-2}\, \Box\phi & = & T \, \gamma \,  e^{\gamma\phi} \ + \ \beta_p \left[e^{-2B} \left({\cal B}_u\right)^2 \ + \  e^{-2D} \left(\nabla_\zeta {\cal B}\right)^2 \right] \, e^{-2\beta_p\phi-2(p+1)A} \nonumber \\
 &+&  h_0^2 \ \beta_{p+1}  \, e^{2 \beta_{p+1}\phi - 2({\cal D}-p-3)C} ,
\eea
where
\bea
\Box\phi &=& e^{-2B}\left( \phi_{uu} \ + \ \phi_u F^+_u\right) \ + \ e^{-2D}\left( \nabla_\zeta \nabla_\zeta \phi \ + \   \nabla_\zeta \phi \nabla_\zeta F^-  \ - \ E_u \nabla_\zeta \phi \right) \ ,
\eea
and finally the equation of motion of the ${\cal B}$ form field is given in eq.~\eqref{eq:Beq}, and we repeat it here for the convenience of the reader
\beq
\nabla_\zeta\left( e^{2G} \ \nabla_\zeta {\cal B} \right)\ = \ E_u \ e^{2G} \ \nabla_\zeta {\cal B} \  - \ \left[e^{2(G -B  + D)} \ {\cal B}_u  \right]_u \ .
\eeq

\subsection{\sc Linear Perturbations of the System}

The $({\cal D}-1)$--dimensional vacuum of~\cite{dm_vacuum} is conveniently captured by the background of eqs.~\eqref{eq:metric_D_dim} with
\bea
A(\zeta,u) &=& B(\zeta) \ , \qquad B(\zeta,u) \ = \ B(\zeta) \ , \qquad C(\zeta,u) \ = \ B(\zeta)  \ + \ \log{u} \ , \nonumber \\
D(\zeta,u) &=& D(\zeta) \qquad E(\zeta,u) \ = \ 0 \ , \qquad  \phi(\zeta,u) \ = \ \phi(\zeta) \ ,
\eea
and with vanishing tensor profiles. The equations of motion satisfied by the background are
\bea
\mu\nu&:&   - \ ({\cal D}-1)\,B_\zeta^2 \ -\ 2 B_{\zeta\zeta}\ +\ 2B_\zeta \, D_\zeta  \ = \  \frac{4}{({\cal D}-2)^2}\ \phi_\zeta^2 \  + \  \frac{T}{{\cal D}-2} \, e^{\gamma\,\phi+2D} \ , \nonumber \\
\zeta\zeta &:&  \frac{4}{{\cal D}-2}\ \phi_\zeta^2-({\cal D}-2)({\cal D}-1) B_\zeta^2 \ = \  T \, e^{\gamma\,\phi+2D} \ , \\
\phi &:& \phi_{\zeta\zeta}  \ + \  \phi_\zeta \left[({\cal D}-1)B_\zeta \ - \  D_\zeta\right] \ = \ \frac{{\cal D}-2}{8} \, T \, \gamma \,  e^{\gamma\phi+2D} \ . \nonumber
\eea
A residual gauge freedom allows reparametrizations of $\zeta$, and can be used to adjust $D$. Two useful gauge choices are $\gamma\phi+2D=0$ and $B=D$: the former allows to write the vacuum in closed form~\cite{dm_vacuum}, while the latter is a conformal gauge, which simplifies somewhat the structure of the equations, at the cost of making the solution implicit.

One can now expand around the vacuum solution, letting
\bea
A &=& B(\zeta)  \ + \ a(\zeta,u) \ , \qquad B \ = \ B(\zeta) \ + \ b(\zeta,u) \ , \qquad C \ = \ B(\zeta) \ + \ \log{u}\  + \  c(\zeta,u) \ ,\nonumber \\
D&=&D(\zeta) \ + \ d(\zeta,u) \ , \qquad \phi \ = \ \phi(\zeta) \ + \ \varphi(\zeta,u) \ , \qquad {\cal B} \ = \ {\cal B}(\zeta,u) \ .
\eea

The $\mu\nu$ components of the linearized metric equations lead to
\bea
&& e^{2(D -B )} \Bigg\{ 2\frac{({\cal D}-p-3)({\cal D}-p-4)}{u^2}(b-c) - 2p a_{uu}-2({\cal D}-p-3)c_{uu} - 2 d_{uu}\nonumber\\
    &&- \ 2\frac{{\cal D}-p-3}{u}\left[p a_u - b_u + ({\cal D}-p-2)c_u + d_u\right] \Bigg\}  - 2 \Bigg[p a_\zeta + b_\zeta + ({\cal D}-p-3)c_\zeta \nonumber \\
        &&- \ \frac{{\cal D}-p-3}{u}E\Bigg]_\zeta+ 2 \left[{D_{\zeta} } - ({\cal D}-1){B }_{\zeta}\right]\left[p a_\zeta + b_\zeta + ({\cal D}-p-3)c_\zeta - \frac{{\cal D}-p-3}{u}E\right] \nonumber \\
            &&+ \ 2 ({\cal D}-2){B }_\zeta (d_\zeta+E_u)+ 2 E_{u\zeta}-2 {D }_\zeta E_u
            \ = \  \frac{8}{{\cal D}-2}{\phi }_\zeta \varphi_\zeta \,  + \,  T \, e^{\gamma\,\phi +2D }(\gamma\varphi + 2 d)   ,
\eea
while the $uu$ component is
\bea
&& e^{2(D -B )} \Bigg[-2 \frac{{\cal D}-p-3}{u} \left[(p+1)a+({\cal D}-p-4)c + d\right]_u \Bigg] \nonumber \\
    && + \ e^{2(D -B )} \, 2\frac{({\cal D}-p-3)({\cal D}-p-4)}{u^2}(b-c)\, - \, 2 \left[(p+1)a_\zeta + ({\cal D}-p-3)c_\zeta - \frac{{\cal D}-p-3}{u}E\right]_\zeta \nonumber \\
        && + \ 2 ({\cal D}-2){B }_\zeta d_\zeta \ + \  2 \left[{D }_\zeta - ({\cal D}-1){B }_\zeta\right]\Bigg[(p+1)a_\zeta+({\cal D}-p-3)c_\zeta- \frac{{\cal D}-p-3}{u}E\Bigg] \nonumber \\
            && = \ \frac{8}{{\cal D}-2}{\phi }_\zeta \varphi_\zeta + T \, e^{\gamma\,\phi +2D } (\gamma\varphi+2 d) \ ,
\eea
the $u\zeta$ component is
\bea
&& e^{2(D -B )}\Bigg[-2(p+1)a_{u \zeta} -2({\cal D}-p-3)c_{u\zeta} + d_u 2({\cal D}-2){B }_{\zeta}+ \frac{2({\cal D}-p-3)}{u}(b_\zeta-c_\zeta)\Bigg]  \nonumber \\
    && + \ ({\cal D}-2) E \ \Big[2{B }_{\zeta\zeta}+({\cal D}-1){B }_\zeta^2-2{B }_\zeta {D }_\zeta  \Big]\nonumber \\
        && = \ e^{2(D -B )} \frac{8}{{\cal D}-2}\ \varphi_u{\phi }_\zeta  -\frac{4E}{{\cal D}-2} \   {\phi }_\zeta^2- T \,E  \, e^{\gamma\,\phi +2D } \ ,
\eea
the $\zeta\zeta$ component is
\bea
&&  -2({\cal D}-2){B }_{\zeta}\left[(p+1)a_\zeta+ b_\zeta + ({\cal D}-p-3)\left(c_\zeta - \frac{E}{u}\right)- E_u\right]\nonumber \\
        && + \  2 (b-c) e^{2(D -B )} \, \frac{({\cal D}-p-3)({\cal D}-p-4)}{u^2} - e^{2(D -B )}\Bigg\{2(p+1)a_{uu} + 2({\cal D}-p-3)c_{uu}\nonumber \\
        && + \  2 \frac{{\cal D}-p-3}{u} \left[(p+1)a_u-b_u+({\cal D}-p-2)c_u\right]\Bigg\} \nonumber \\
                && = \ -\frac{8}{{\cal D}-2}\ {\phi }_\zeta \varphi_\zeta  \ +  \  T \, e^{\gamma\,\phi +2D } (\gamma \varphi + 2 d) \ ,
\eea
and finally the $mn$ components lead to
\bea
&& 2 (b-c)e^{2(D -B )} \frac{({\cal D}-p-4)({\cal D}-p-5)}{u^2}+ \  e^{2(D -B )}\Bigg\{-2\Big[(p+1)a_{uu}+({\cal D}-p-4)c_{uu} \nonumber \\
    && + \ d_{uu}\Big]- 2 \frac{{\cal D}-p-4}{u}\Big[(p+1)a_u-b_u+({\cal D}-p-3)c_u + d_u\Big] \Bigg\} \nonumber \\
            && - \ 2 \left[(p+1)a_\zeta + b_\zeta + ({\cal D}-p-4)c_\zeta - \frac{{\cal D}-p-4}{u} E\right]_\zeta \nonumber  \\
                && + \ 2\left[D_\zeta - ({\cal D}-1)B_\zeta\right]\left[(p+1)a_\zeta + b_\zeta + ({\cal D}-p-4)\left(c_\zeta - \frac{E}{u}\right)\right]  \nonumber\\
                    && + \ 2 E_u \left[({\cal D}-2)B_\zeta - D_\zeta\right] +2 E_{u\zeta} +2 ({\cal D}-2) B_\zeta d_\zeta\nonumber \\
                        && = \ \frac{8}{{\cal D}-2}\phi_\zeta \varphi_\zeta +T \, e^{\gamma\,\phi+2D} (\gamma\varphi+2d) \ .
\eea
The linearized dilaton equation reads
\bea
&& e^{2(D -B )}\left[ \varphi_{uu}+\varphi_u ({\cal D}-p-3)\frac{1}{u}\right] +  \varphi_{\zeta\zeta} \nonumber \\
&&+ \ {\phi }_\zeta \left[ (p+1)a_\zeta + b_\zeta + ({\cal D}-p-3)c_\zeta - d_\zeta - E \frac{({\cal D}-p-3)}{u} \right]  \nonumber \\
&&+ \  \varphi_\zeta \left[({\cal D}-1){B }_\zeta - {D }_\zeta\right]  - E_u {\phi }_\zeta  \ = \ \frac{{\cal D}-2}{8}T \, \gamma \,  e^{\gamma\phi +2D } \left(\gamma \varphi + 2 d\right) \ ,
\eea
and finally the linearized equation of motion of the form field reads
\beq
{\cal B}_{\zeta\zeta}\ + \ (2G)_\zeta {\cal B}_\zeta  \ = \ - \ e^{2(-B   + D )} \ {\cal B}_{uu} \ - \ e^{2(-B   + D )} \frac{{\cal D}-p-3}{u} \ {\cal B}_{u} \ .
\eeq

The preceding system of equations admits the gauge transformations
\beq
u \ \to \  u \ + \ \delta u(\zeta,u) \ , \qquad \zeta \ \to \  \zeta \ + \ \delta \zeta(\zeta,u) \ ,
\eeq
which translate into the first--order variations
\bea \label{eq:gauge_transformations}
\delta a &=&B_\zeta \,\delta \zeta \ , \qquad \delta b \ = \ B_\zeta\, \delta \zeta + (\delta u)_u \ , \qquad \delta c \ = \ B_\zeta \, \delta \zeta +\frac{\delta u}{u} \ , \qquad
\delta d \ = \  D_\zeta \,\delta \zeta + (\delta \zeta)_\zeta \ , \nonumber \\ \delta E &=& (\delta u)_\zeta + e^{2(D-B)}(\delta \zeta)_u \ , \qquad \delta\varphi \ = \ \phi_\zeta\, \delta\zeta \ , \qquad \delta{\cal B} \ = \ 0 \ .
\eea

It is now convenient to work in the conformal gauge for the background, so that
\beq
B(\zeta) \ = \ D(\zeta) \ = \ \Omega(\zeta) \ ,
\eeq
and we can now see in detail how the residual gauge transformations can be used to set $E=0$ and $b=c$.
To begin with, the transformation in the last line of eqs.~\eqref{eq:gauge_transformations}
\beq
\delta E \ = \ (\delta u)_\zeta \ + \  (\delta \zeta)_u
\eeq
allows one to remove $E$, while leaving some residual gauge freedom, with $\delta u$ and $\delta \zeta$ satisfying the Cauchy--Riemann equations
\beq
(\delta u)_\zeta \ + \ (\delta \zeta)_u \ = \ 0 \ .
\eeq
This reduces the independent gauge parameters, so that one is left with
\beq
\delta u \ =\  f_u \ , \qquad \delta \zeta \ = \ - \ f_\zeta \ ,
\eeq
where $f(\zeta,u)$ is an arbitrary function of $\zeta$ and $u$.
Nonetheless, one can still choose the gauge $b=c$, since $b-c$ transforms as
\beq
\delta(b-c) \ = \  f_{uu} \ - \ \frac{1}{u} \, f_u \ .
\eeq
In principle, there are even residual gauge transformations, which satisfy
\beq
f_{uu} \ - \ \frac{1}{u} \, f_u \ = \ 0 \ ,
\eeq
and are thus of the form
\beq
f \ = \ u^2 \, f_1(\zeta) \ + \ f_2(\zeta) \ .
\eeq
However, $f_1(\zeta)$ and $f_2(\zeta)$ must vanish, since they would affect the limiting behavior of the perturbations, which are supposed to wane at infinity. Abiding to the notation of the main body of the paper, we now call $z$ and $r$ the gauge fixed $(\zeta,u)$ variables.

In the gauge $E=0$, $b=c$, the linearized system reduces to
\bea
&-&   2 \left[ p a + ({\cal D}-p-3)c  + d \right]_{rr} - 2\frac{{\cal D}-p-3}{r}\left[p a + ({\cal D}-p-3)c + d\right]_r   \nonumber\\
        &-& 2 \left[p a  + ({\cal D}-p-2)c \right]_{zz} - 2 ({\cal D}-2){\Omega}_z\left[p a + ({\cal D}-p-2)c - d \right]_z \nonumber \\
            &=&  \frac{8}{{\cal D}-2}\,{\phi }_z \varphi_z \  + \  T \, e^{\gamma\,\phi +2\Omega }(\gamma\varphi + 2 d) \ ,\\
&-&\left[ p a +({\cal D}-p-3)c  + d \right]_{rr} + \frac{{\cal D}-p-3}{r}\left(a -c \right)_r  +\left(a - c \right)_{zz} \nonumber\\
        &+& ({\cal D}-2){\Omega }_z\left( a - c \right)_z  = 0 \ , \\
&-&2({\cal D}-2){\Omega }_{z}\left[(p+1)a_z + ({\cal D}-p-2)c_z\right]-2 \left[(p+1)a+({\cal D}-p-3)c\right]_{rr}\nonumber \\
        &-&   2 \frac{{\cal D}-p-3}{r} \left[(p+1)a+({\cal D}-p-3)c\right]_r \nonumber \\
                &=& -\frac{8}{{\cal D}-2}\ {\phi }_z \varphi_z  \ +  \  T \, e^{\gamma\,\phi +2\Omega } (\gamma \varphi + 2 d) \ , \\
&-&2\left[(p+1)a+({\cal D}-p-4)c +d \right]_{rr}-2 \frac{{\cal D}-p-4}{r}\left[(p+1)a +({\cal D}-p-4)c+ d\right]_r  \nonumber \\
            &-& 2 \left[(p+1)a  + ({\cal D}-p-3)c\right]_{zz} -2({\cal D}-2)\Omega_z\left[(p+1)a+({\cal D}-p-3)c -d \right]_z  \nonumber\\
                    &=& \frac{8}{{\cal D}-2}\phi_z \varphi_z +T \, e^{\gamma\,\phi+2\Omega} (\gamma\varphi+2d) \ ,
\eea
together with the $rz$ equation, which can now be integrated in $r$, taking into account that perturbations are to wane at infinity, to give
\beq
- \ 2(p+1)a_{z}  \ - \ 2({\cal D}-p-3)c_{z} \ + \  2({\cal D}-2) \,{\Omega }_{z} \, d \ = \ \frac{8}{{\cal D}-2}\ \varphi\,{\phi }_z  \ .
    \eeq
This determines the dilaton perturbation $\varphi$ in terms of the metric ones, and one can show that the dilaton equation, which becomes
\bea
& &  \varphi_{rr}+ \frac{{\cal D}-p-3}{r}\,\varphi_r +  \varphi_{zz} +{\phi }_z \Big[ (p+1)a + ({\cal D}-p-2)c - d  \Big]_z  \nonumber \\
&+&  ({\cal D}-2)\varphi_z \,{\Omega }_z \ = \ \frac{{\cal D}-2}{8}T \, \gamma \,  e^{\gamma\phi +2\Omega } \left(\gamma \varphi + 2 d\right) \ ,
\eea
follows from the metric ones, so that we are leaving it aside in the main body of the paper.
Finally the linearized equation of motion of the form field reduces to
\beq
{\cal B}_{zz}+ (2G)_z\, {\cal B}_z  \ = \ - \ {\cal B}_{rr} -  \frac{{\cal D}-p-3}{r} \ {\cal B}_{r} \ .
\eeq

The following steps are carried out in detail in Section~\ref{sec:linearized_analysis} for ${\cal D}=10$, the case of direct interest for String Theory. However, the system for generic values of ${\cal D}$ can prove useful in the discussion of lower--dimensional models of this type.

\newpage

\end{appendices}

\end{document}